\documentclass[smallextended]{svjour3} 

\usepackage{graphicx}
\usepackage{orcidlink}
\usepackage{enumitem}

\usepackage{url}
\usepackage{tabularx}
\usepackage{booktabs}
\usepackage{colortbl}

\usepackage{framed}
\usepackage{multirow}
\usepackage{natbib}

\definecolor{SkyBlue}{rgb}{0.53, 0.81, 0.92} 
\usepackage[framemethod=TikZ]{mdframed}
\usepackage{hyperref}

\definecolor{light-gray}{gray}{0.9}

\hypersetup{
    colorlinks=true,
    linkcolor=black,
    filecolor=magenta,      
    urlcolor=blue,
    citecolor=black,
    pdfborder={0 0 0}
}

\mdfdefinestyle{MyFrame}{%
    linecolor=black,
    outerlinewidth=0.15pt,
    roundcorner=3pt,
    innertopmargin=2pt,
    innerbottommargin=2pt,
    innerrightmargin=4pt,
    innerleftmargin=4pt,
    backgroundcolor=light-gray}
    
\mdfdefinestyle{MyFrameSimple}{%
    linecolor=black,
    outerlinewidth=0.15pt,
    roundcorner=3pt,
    innertopmargin=2pt,
    innerbottommargin=2pt,
    innerrightmargin=4pt,
    innerleftmargin=4pt}

\newcommand{\sectopic}[1]{\vspace*{0.1em}\par\noindent{\textit{\bfseries #1}}}

\journalname{Empirical Software Engineering}

\begin{document}

\title{An Empirical Study on LLM-based Classification of Requirements-related Provisions in Food-safety Regulations}

\titlerunning{Classification of Requirements-related Provisions in Food-safety Regulations}

\author{{Shabnam Hassani \textsuperscript{1} \orcidlink{0009-0008-3056-4073}, Mehrdad Sabetzadeh \textsuperscript{1} \orcidlink{0000-0002-4711-8319}, and Daniel Amyot \textsuperscript{1} \orcidlink{0000-0003-2414-1791}}}

\authorrunning{S. Hassani, M. Sabetzadeh, and D. Amyot}

\institute{Shabnam Hassani\at
              \email{s.hassani@uottawa.ca}
           \and
           Mehrdad Sabetzadeh \at
              \email{ m.sabetzadeh@uottawa.ca} 
           \and
           Daniel Amyot  \at
              \email{ damyot@uottawa.ca}      
\\
\textsuperscript{1} University of Ottawa, Ottawa, ON, Canada
}

\date{Received: date / Accepted: date}

\maketitle

\begin{abstract}
As Industry 4.0 transforms the food industry, the role of software in achieving compliance with food-safety regulations is becoming increasingly critical. Food-safety regulations, like those in many legal domains, have largely been articulated in a technology-independent manner to ensure their longevity and broad applicability. However, this approach leaves a gap between the regulations and the modern systems and software increasingly used to implement them. In this article, we pursue two main goals. First, we conduct a Grounded Theory study of food-safety regulations and develop a conceptual characterization of food-safety concepts that closely relate to systems and software requirements. Second, we examine the effectiveness of two families of large language models (LLMs) -- BERT and GPT -- in automatically classifying legal provisions based on requirements-related food-safety concepts. 
Our results show that: (a) when fine-tuned, the accuracy differences between the best-performing models in the BERT and GPT families are relatively small. Nevertheless, the most powerful model in our experiments, GPT-4o, still achieves the highest accuracy, with an average \emph{Precision} of 89\% and an average \emph{Recall} of 87\%; (b) few-shot learning with GPT-4o increases \emph{Recall} to 97\% but decreases \emph{Precision} to 65\%, suggesting a trade-off between fine-tuning and few-shot learning; (c) despite our training examples being drawn exclusively from Canadian regulations, LLM-based classification performs consistently well on test provisions from the US, indicating a degree of generalizability across regulatory jurisdictions; and (d) for our classification task, LLMs significantly outperform simpler baselines constructed using long short-term memory (LSTM) networks and automatic keyword extraction.
\end{abstract}

\section{Introduction} \label{sec:introduction}
Software is increasingly pervasive in regulated industries where compliance with legal provisions is a must. Examples of regulated industries include healthcare, transportation, energy, food and agriculture. Driven by pressing concerns such as data protection and privacy, some industries, such as healthcare, have incorporated specific measures into their compliance frameworks to better address the role of software~\citep{Breaux2006Towards}. Nonetheless, many industries, where the pressure to adapt has not been as strong, have yet to give adequate consideration to software. Our work in this article is focused on one of these industries, namely the \emph{food industry}.

The most critical regulatory aspect for food is \emph{food safety}. Food safety refers to the measures taken to ensure that food products are free from harmful contaminants and that they are safe for human consumption~\citep{WHO2024FoodSafety}. 
Food-safety regulations set standards and guidelines for the production, processing, packaging, labelling, storage, and distribution of food products. 
Food safety is arguably the most far-reaching safety concern, as it affects the life of every individual. This makes food-safety regulations among the most impactful and significant in terms of their importance. An estimated 600 million people -- nearly 1 in 13 -- fall ill each year from contaminated food, with 420,000 fatalities, disproportionately affecting children under five. Economically, unsafe food results in an estimated annual loss of \$110 billion USD in productivity and medical expenses in low- and middle-income countries, placing a significant strain on healthcare systems and impeding socioeconomic development~\citep{WHO2024FoodSafety}. To address food-safety risks, there is a robust and growing market for food-safety testing and monitoring, the size of which is projected to be around \$24 billion USD by the end of 2024~\citep{FoodSafetyTesting}.

In recent years, Industry 4.0~\citep{Gilchrist2016Industry4.0} has significantly transformed the food industry. One of the main thrusts of this transformation has been the introduction of the Internet of Things (IoT) into food supply chains. This has enabled companies to monitor and control various processes, including those related to food safety, in real-time \citep{Bouzembrak2019Internet}.
For example, food-service operators can now use sensors to monitor environmental conditions such as temperature and humidity during food storage and transportation to ensure that food products are maintained at optimal conditions.

Food-safety regulations, like many other types of legal measures, are \linebreak technology-agnostic and are designed to ensure long-term relevance, promote innovation, and maintain market neutrality. While this flexibility is beneficial, it also creates a gap between regulations and modern food-safety systems, which increasingly rely on software~\citep{menard2024food,gottschald2024advancing}. Closing this gap requires developing techniques that enable a smooth transition from regulations to systems and software requirements.

\vspace*{.5em}\par\noindent\textbf{Motivation.} Our research was prompted by a stated need from a collaborating industry partner, Stratosfy (\url{https://stratosfy.io}). Stratosfy develops an IoT-based temperature monitoring system for food products. This system helps food businesses such as restaurants and supermarkets avoid spoilage by monitoring their cooling devices (e.g., walk-in coolers, reach-in coolers and milk dispensers), and sending warnings if anomalies happen in temperature patterns. 
A frequent request from the clients of this system is to provide additional features that would facilitate demonstration of compliance with food-safety regulations. The partner is thus interested in transforming food-safety provisions into feature requirements in line with the capabilities afforded by the existing IoT technologies. 

Identifying which food-safety provisions are relevant to systems and software requirements is not a straightforward task, since food-safety regulations have not been written with systems and software in mind. The \emph{first challenge} that arises here is characterizing the food-safety concepts that are most relevant to automated food-safety systems.

A \emph{second challenge} is posed by the sheer volume of text in food-safety regulations. Even with a solid understanding of the relevant concepts and expertise in legal language, manually searching for instances of these concepts within vast amounts of legal content -- often across multiple jurisdictions -- can be an overwhelming task. A prime example of the scale involved is the Codex Alimentarius (Latin for ``Food Code''), developed jointly by the Food and Agriculture Organization (FAO) and World Health Organization (WHO). Established in 1963, the Codex is an intentionally recognized compilation of standards, guidelines, and codes of practice aimed at ensuring the safety, quality, and fairness of international food trade. As of this writing, the Codex has 87 guidelines, 237 commodity standards, and 57 codes of practice, totaling \emph{thousands} of pages on material related to food safety. Individual countries adapt this material into their national regulatory frameworks, often tailoring them to address local public health concerns, consumer preferences, and trade priorities. This process frequently results in the creation of extensive regulatory corpora, which can amount to thousands of pages for each country~\citep{CodesofPractice,Guidelines,Standards}. When dealing with a large volume of legal documents, automated assistance can substantially improve efficiency and accuracy in identifying provisions that are requirements-related.

\begin{figure}[!t]
\centering
\includegraphics[width=.93\linewidth] {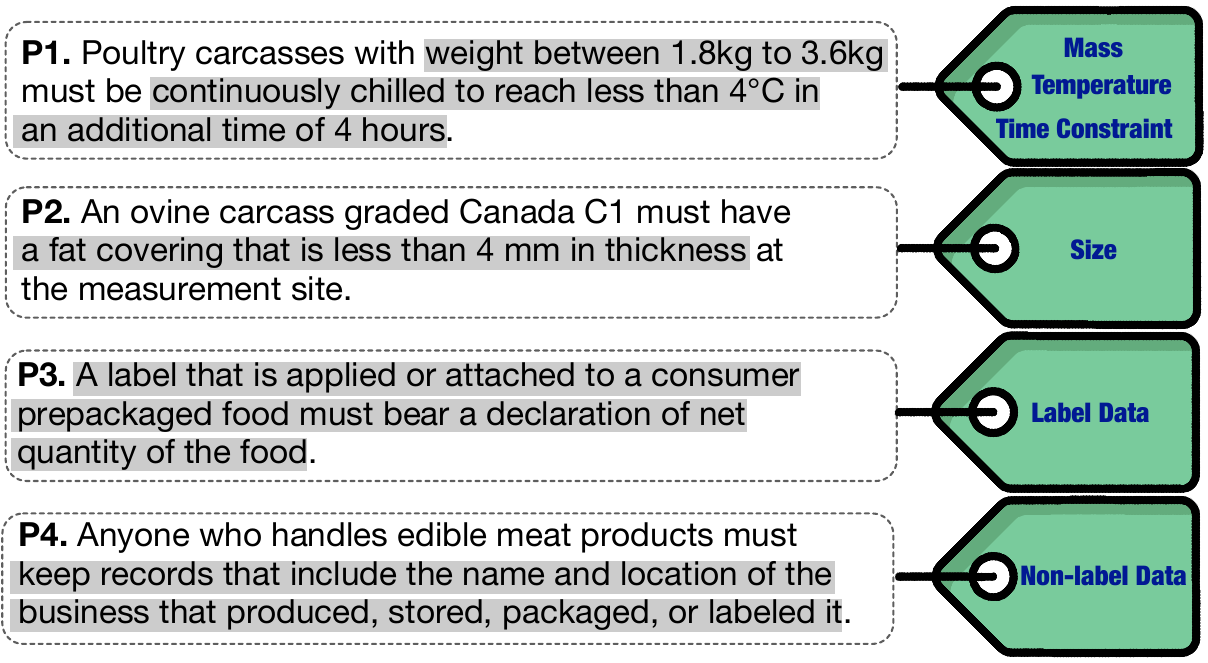}
\caption{Example provisions, \textsf{P1}--\textsf{P4}, from the Safe Food for Canadians Regulations \citep{SFCR2018} alongside labels that classify the requirements-related content of these provisions.} \label{fig:example}
\end{figure}

\vspace*{.5em}\par\noindent\textbf{Objectives.} The objectives of this article are two-fold: {(1)}~to provide a characterization of regulatory concepts that are relevant to the development of food-safety systems, and {(2)}~to develop an automated pipeline to accurately identify instances of these concepts in an input legal text.

We illustrate our objectives using the example of Fig.~\ref{fig:example}. The provisions \textsf{P1}--\textsf{P4} in this figure originate from the Safe Food for Canadians Regulations (SFCR) \citep{SFCR2018}, discussed further in Section~\ref{sec:Back}. These have been modified from their original form to simplify illustration.
The original provisions are available in our online material~\citep{Data}.

In our example, \textsf{P1} could imply sensing requirements for the measurement of \emph{Mass} and \emph{Temperature}. Furthermore, \textsf{P1} is time-sensitive, indicating a \emph{Time Constraint} that can potentially impact monitoring requirements. \textsf{P2} may imply additional sensing requirements, say if digital calipers are used to measure fat thickness (\emph{Size}). \textsf{P3} specifies information that food \textit{Label}s must include, with potential implications for sensing, monitoring and data management. \textsf{P4} envisages important traceability information \emph{beyond} food-product labels (i.e., \emph{Non-label Data}) \hbox{that may need to be collected during operation.}

Although none of the provisions in Fig.~\ref{fig:example} directly refer to automation, the provisions can induce different systems and software requirements depending on the nature and extent of automation implemented in a proposed food-safety system. 
Our first objective has to do with defining what labels (classes), e.g., \emph{Temperature} and \emph{Time Constraint}, to classify provisions by. Given a food-safety-related legal text, our second objective is concerned with automatically attaching labels to the provisions in the text, as exemplified in Fig.~\ref{fig:example}. We are particularly interested in \emph{Large Language Models (LLMs)} as enablers of such automation. These models are capable of understanding, generating, and interpreting natural language in context, thus offering the potential to more accurately identify and classify relevant regulatory concepts.

{\vspace*{.5em}\par\noindent\textbf{Research Questions (RQs):}\label{subsec:RQs}}

We achieve our objectives by answering the following RQs: 

\begin{itemize}
\item\textbf{RQ1: What concepts in food-safety regulations are most relevant to systems and software requirements?} We answer RQ1 by conducting a Grounded Theory (GT) study, focusing on Canadian food-safety regulations.

\item\textbf{RQ2: How accurately can LLMs classify requirements-related provisions in food-safety regulations?} We answer RQ2 by systematically examining classification models built using variants from two LLM families: BERT and GPT.

\item \textbf{RQ3: How do LLM-based classification models compare to simpler baselines in the context of food-safety regulations?} We answer RQ3 by comparing the best-performing LLM-based classification models identified in RQ2 with two baseline methods: one using a long short-term memory (LSTM) model and the other based on keyword search.

\end{itemize}

\vspace*{.5em}\par\noindent\textbf{Contributions.}\label{subsec:contribution} The contributions of this article are as follows: 
\begin{enumerate}[leftmargin=*]
    \item \emph{A Grounded Theory (GT) study of Canadian food-safety regulations.}
    Using a GT study, we  analyze federal food-safety regulations in Canada to identify key concepts that influence food-safety system requirements. We focus on Canadian regulations because they are well-established and closely aligned with the operational domain of our industry partner.
    
    \item \emph{An empirical evaluation of LLMs for classifying food-safety provisions.} We create an automated pipeline that can be instantiated with different LLMs for classifying the content of legal texts related to food safety. We use the information elements from the content model developed in our GT study (i.e., our first contribution) as classification labels. We conduct detailed experimentation with state-of-the-art variants (as of this writing) from two families of LLMs: BERT and GPT.  
    Our results show that: (a) when fine-tuned, the accuracy differences between the best-performing models in the BERT and GPT families are relatively small. Nevertheless, the most powerful model in our experiments, GPT-4o, still has an edge with an average \emph{Precision} of 89\% and an average \emph{Recall} of 87\%; (b) few-shot learning with GPT-4o increases \emph{Recall} to 97\% but decreases \emph{Precision} to 65\%, suggesting a trade-off between fine-tuning and few-shot learning; (c) despite our training examples being drawn exclusively from Canadian regulations, LLM-based classification performs consistently well on test provisions from the US, indicating a degree of generalizability across regulatory jurisdictions; and (d) for our classification task, LLMs significantly outperform simpler baselines constructed using long short-term memory (LSTM) networks and automatic keyword extraction.
\end{enumerate}
\vspace*{.5em}\par\noindent\textbf{Target Users and Practical Applications.}\label{practical_usecases}
Our work aims to establish a conceptual basis for the development of regulatory requirements for food-safety systems and to systematically assess the readiness of LLMs as an assistive technology in this context. While acknowledging that our conceptualization of requirements-related concepts in food-safety regulations and our automated support for identifying these concepts have not yet been validated in industrial applications, our contributions target the following stakeholders as the main beneficiaries: suppliers and integrators of food-safety monitoring systems, who are responsible for ensuring compliance with relevant regulations. To assist these stakeholders, our work paves the way for implementing two important use cases:

\begin{enumerate}
\item \textbf{Use Case 1 (Filtering Irrelevant Content)}. One major advantage of automated classification is providing the ability to filter out irrelevant content. Our GT study indicates that only about 25\% of the paragraphs in SFCR documents are relevant, meaning that they include occurrences of requirements-related concepts. By developing an accurate automated content classifier, we provide an effective tool to filter out content unrelated to systems and software requirements. This in turn enables limited human resources to focus on content that is more likely to be relevant and have a significant impact on requirements.

\item \textbf{Use Case 2 (Metadata-driven Guidance)}. Regulations often contain complex legal jargon, increasing the likelihood that requirements-relevant content may be overlooked if reviewed without guidance. To address this issue, an additional layer of \emph{metadata} is useful for highlighting how specific provisions relate to systems and software, such as by requiring some kind of sensing, measurement, or data collection. As shown by our illustrative example in Fig.~\ref{fig:example}, classification labels help clarify the nature of each provision's relevance to systems and software requirements. This in turn enables stakeholders to focus more effectively on the relevant aspects of the regulations.
\end{enumerate}

\vspace*{.5em}\par\noindent\textbf{Novelty.} Compliance with legal provisions is a key concern for software-intensive systems. This has led the Requirements Engineering (RE) community to develop automation for classifying and extracting requirements-related information from legal texts. However, most previous research in this direction, e.g., from \cite{Zeni2015Gaiust,Zeni2016Building} and \cite{Sleimi2018Automated}, has focused on high-level concepts such as those in the Hohfeldian system \citep{Hohfeld2008Fundamental} or deontic logic \citep{Jones1992Deontic}. While such research is useful for domains where a clear conceptual link exists between law and software-intensive systems, it is insufficient for domains like food safety, where no such link has been established yet. The RE literature further acknowledges the importance of domain-focused classification of legal concepts \citep{Breaux2006Towards,Ghanavati2007Towards}. Yet, most current studies address privacy regulations \citep{Evans2017Evaluation,Bhatia2018Semantic,Torre2020Ai,Torre2021Modeling,Xie2022Scrutinizing,Amaral2022AI}, which are relatively new and align closely with software systems. The study of traditional domains whose regulatory frameworks were not designed with such systems in mind remains limited. Our work is the first to attempt to address this gap for the domain of food safety.

\vspace*{.5em}\par\noindent\textbf{Structure.} Section~\ref{sec:Back} discusses background. Section~\ref{sec:metadata} presents our GT study. Section~\ref{sec:extraction} describes our pipeline for automated text processing and classification. Section~\ref{sec:evaluation} reports on the accuracy of this pipeline for food-safety regulations, experimenting with variants from the BERT and GPT LLM families. Section~\ref{sec:discussion} outlines key practical implications. Section~\ref{sec:related} compares with related work. Section~\ref{sec:conclusion} provides concluding remarks. Section~\ref{sec:package} directs to our replication package.

\section{Background} \label{sec:Back}
This section presents the necessary background on food-safety regulations and on the enabling AI technologies used.  
\subsection{Food-safety Regulations}\label{sec:FoodRegulations} 
To ensure that the food supply is safe, countries around the world have established regulatory frameworks to govern the production, distribution, and consumption of food products \citep{WHO2024FoodSafety}. 
Our primary focus is on the Safe Food for Canadians Regulations (SFCR) \citep{SFCR2018}.
SFCR consolidates several federal Canadian food regulations that apply, among other things, to the import, export, and inter-provincial trade of food, including ingredients \citep{CFIA2018Understanding}.
SFCR sets out both general food requirements as well as specific requirements for individual food products.
In many cases, SFCR defers the elaboration of requirements for specific food products to Food-Specific Requirements and Guidance (FSRG) regulations \citep{FSRG2018}. The FSRG regulations notably cover dairy products, egg products, fish, fruits and vegetables, meat, honey, manufactured products, and maple. We use content from FSRG regulations both in our GT study (Section~\ref{sec:metadata}) and in our evaluation (Section~\ref{sec:evaluation}).

In addition to SFCR and the associated FSRG regulations, we consider in our work selected parts of the food-safety regulations by the US Food and Drug Administration (FDA) \citep{FDA1906}. FDA regulates various aspects of food production and handling, including labelling, ingredients and packaging. We use FDA regulations to examine the extent to which our classification solution is transferable \hbox{to jurisdictions outside Canada.}

\subsection{Grounded Theory (GT) \label{subsec:GT}} 
To identify food-safety concepts that are relevant to systems and software requirements, we apply GT. GT is a qualitative research method originally described by~\cite{glaser1967discovery}. The primary objective of GT is to develop theory through the systematic collection and analysis of data, rather than to test or validate pre-existing theories. Noting that, to the best of our knowledge, no systematic conceptualization (theory) of food-safety concepts related to systems and software requirements has been previously established, GT offers an effective method for deriving such a conceptualization in a methodical and well-documented way.

Over time, GT has evolved into several variants, most notably Glaserian GT (developed by \citet{Glaser1992Basics}), Straussian GT (developed by ~\citet{Strauss1998Basics}), and Constructivist GT (developed by \citet{Charmaz2014Constructing}). \citet{Stol2016Grounded} explore the commonalities and differences among the GT variants from a software-engineering perspective. In this article, we employ Straussian GT to analyze Canadian food-safety regulations. We select this variant for two primary reasons: (1) our study addresses a well-defined research question (RQ1, as presented in Section~\ref{sec:introduction}), and (2) as discussed further in Section~\ref{sec:metadata}, throughout our study, we require the flexibility to engage with the existing literature, which Straussian GT permits.

Straussian GT has the following phases~\citep{Strauss1998Basics, Stol2016Grounded, Moshtari2022Grounded}:
\textbf{Defining Research Questions:} Researchers start with broad, open-ended questions that guide the investigation. \textbf{Theoretical Sampling:} In this phase, researchers select data sources based on their ability to contribute to the emerging theory. Sampling continues until theoretical saturation is achieved, meaning no new insights or concepts emerge. \textbf{Open Coding:} This phase involves breaking down the data into discrete parts to identify concepts. Labels are assigned to concepts, and emergent categories are recognized. \textbf{Constant Comparisons:} Throughout the process, data is continuously compared to refine categories, identify relationships, and ensure no gaps in the theory. \textbf{Memoing:} Researchers document their preliminary ideas about the properties and conceptual relationships between categories through memos, which can include notes, diagrams, or sketches. \textbf{Axial Coding:} After establishing categories, axial coding links categories and subcategories, enabling a deeper understanding of their interrelationships. \textbf{Selective Coding:} Finally, a core category is identified, and the various categories are integrated into a coherent theory that addresses the research question.

All variants of GT, including Straussian GT, are inherently iterative, meaning that one may revisit earlier phases as new insights emerge. For example, in Straussian GT, one might return to open coding or adjust theoretical sampling based on discoveries made during axial or selective coding. This non-linear approach also allows different phases to occur in parallel, such as memoing throughout the entire process or engaging in axial and selective coding simultaneously as the theory takes shape. This flexibility ensures that the resulting theory is properly grounded in the data and remains responsive to the evolving research context.

\subsection{Large Language Models (LLMs)} \label{subsec:LMs}
We use large language models (LLMs) to classify information in food-safety regulations. LLMs are statistical models trained to predict and generate coherent, contextually relevant text.
These models are broadly categorized into generative models, like GPT \citep{Radford2018Improving}, which are optimized for text production, and discriminative models, such as BERT \citep{Devlin2018Bert}, which are optimized for classification and analysis tasks. LLMs are typically pre-trained on large volumes of textual data to learn contextual information, language regularities, and syntactic and semantic relationships. Following pre-training, LLMs usually require \emph{fine-tuning} to achieve optimal performance in downstream tasks \citep{Vaswani2017Attention}. Fine-tuning involves adjusting a (pre-trained) model to tailor it to a specific application or domain.

Generative AI models, including generative LLMs, commonly use \emph{prompts} as input instructions to guide the model, enabling users to specify the desired nature or context of the generated output \citep{hariri2023unlocking}. Various prompting strategies, such as zero-shot, few-shot, chain-of-thought, and tree-of-thought prompting, exist to enhance output quality \citep{Liu2023pre}.

Below, we briefly discuss the LLMs explored in this article.

\textit{Bidirectional Encoder Representations from Transformers (BERT)}\footnote{With the rapid development of generative models, now commonly having billions of parameters, the term ``LLM'' is evolving and is increasingly used interchangeably with generative language models. Although BERT was not designed for text generation, its architecture, scale, and linguistic capabilities share significant similarities with modern generative models like GPT. Therefore, in this article, we include BERT under the term LLM.} employs bidirectional training of transformers -- a popular attention model for language modelling \citep{Devlin2018Bert}. BERT comes in two sizes: \emph{base} and \emph{large}. While BERT large has better accuracy in certain tasks, it is computationally more expensive. BERT can handle capitalization distinctions; the \emph{uncased} version converts text to lowercase before tokenization, while the \emph{cased} version preserves capitalization.
Prior research favours the cased model for analyzing requirements and regulatory documents \citep{Hey2020NoRBERT, Luitel2023Using}. We therefore adopt the cased model. BERT has several variants that aim to enhance the performance of the original BERT model. In addition to BERT base and BERT large, we experiment with the following BERT variants in this article: \textit{ALBERT} \citep{Lan2019Albert} 
and \textit{RoBERTa} \citep{Liu2019Roberta}. 

\textit{Generative Pre-training Transformer (GPT)} is a transformer-based family of models, pre-trained to predict the next token in a sequence \citep{Radford2018Improving}. GPT has shown the ability to capture subtle linguistic patterns and perform well in many language tasks \citep{Brown2020Language, GPT42023Technical}. In this article, we use GPT-3.5-turbo and GPT-4o for experimentation. These are our best choices as of this writing: The models are both recent and fine-tunable at costs that would allow for the extensive empirical examination necessitated by our experimental procedure.

We consider two other LLM families -- Llama and Mixtral -- in addition to BERT and GPT. These are introduced below but are ultimately excluded from our  evaluation due to technical limitations, as detailed in Section~\ref{subsec:RQ2}.

\textit{Large Language Model Meta AI (Llama)} is a group of open-weight LLMs developed by Meta~\citep{touvron2023llama},~\citep{Llama} for different natural language processing tasks. In this article, we use Llama-3-8B-Instruct, a model specifically fine-tuned for following instructions.

\textit{Mixtral} is the first open-weight sparse mixture-of-experts (sMOE) model developed by Mistral AI~\citep{jiang2024mixtral,Mixtral}. By activating only the relevant parameters based on the task, Mixtral optimizes computational efficiency while maintaining high accuracy. In this article, we explore Mixtral-8x7B-Instruct-v0.1, a variant specifically fine-tuned for instructional tasks.

\subsection{Long Short-Term Memory (LSTM)} 
An alternative model for classification is long short-term memory (LSTM) -- a type of recurrent neural network designed to accurately process long sequences of data \citep{Hochreiter1997Long}. Compared to LLMs, LSTM models have fewer parameters and do not require as much computational power to train and run. LSTM is not a part of our solution. Rather, we use LSTM as a baseline to assess whether employing the more computationally expensive LLMs is justified for our intended application. We build our LSTM baseline using BiLSTM. BiLSTM uses information from both past and future elements in an input sequence by processing the sequence in both forward and backward directions using two separate LSTM layers \citep{Graves2005Framewise}.

\section{Characterization of Requirements-related Information in Food-safety Provisions} \label{sec:metadata}
In this section, we address RQ1 (as posed in Section~\ref{sec:introduction}) by conducting a GT study, following the Straussian variant of GT as explained in Section~\ref{subsec:GT}.
The GT study was conducted collaboratively by the first two authors: a PhD student with two years of experience in qualitative and GT analysis, and a senior researcher with 12 years of experience in these areas. The results were subsequently validated by the last author.

\sectopic{Study Context and Data Selection.} 
We base our GT study on food-safety regulations in Canada. This decision is made in view of two factors: 
(1)~The majority of our industry partner's market is in Canada, 
and (2)~Canada has one of the most comprehensive regulatory frameworks for food safety, evidenced by the country's top ``Quality and Safety'' ranking in the Global Food Security Index (GFSI) \citep{GFSI}. While variations are to be expected in other countries and jurisdictions, we anticipate that the concepts identified in our study should generalize to food-safety systems elsewhere.

In our GT study, we analyze the textual content of SFCR and the FSRG regulations introduced in Section~\ref{sec:FoodRegulations}. 
Our data collection follows a theoretical sampling approach, where we carefully choose additional food-safety provisions to analyze based on emerging concepts and patterns, aiming to quickly reach saturation.
Quick saturation minimizes unnecessary data collection, saving time and effort, while also improving analysis by allowing researchers to focus on richer data that has the potential to yield deeper insights~\citep{Stol2016Grounded}.

To achieve rapid saturation, we prioritize selecting FSRG regulations that we find more likely to refine or complement SFCR. To this end, we choose the regulations for meat and egg products. These products are inherently more complex to prepare, process, maintain, and trade. As a result, they strike a good balance between managing the resource-intensive GT analysis and increasing the likelihood of identifying the most relevant concepts for food-safety systems.
For example, a recurring theme in meat regulations is the need for precise and continuous mass measurement. This theme emerged as a salient concept, with meat being a very rich category of food products in which such measurements are taken. A thorough examination of mass measurement for meat products thus reduced the need to repeatedly assess similar content for other food products.

Specifically, a selected subset of the provisions in these regulations was coded by third-parties, as we explain in Section 5.2. The main distinction is that the authors applied a Grounded Theory (GT) approach. In contrast, the third-parties applied hypothesis coding using the author-defined codes as the basis for their work. No conflicts or new codes emerged from the follow-on coding by third-parties.

\sectopic{Analysis Procedure.} 
We reviewed all the sentences in the SFCR, followed by the FSRG regulations on meat and egg. The goal of this review is to identify any sentences that could potentially have some bearing on systems and software requirements. SFCR and the FSRG regulations have cross-references to other documents, e.g., Safe Food for Canadians Act \citep{SFCA2012} and Food and Drug Act \citep{FDA1985}. We examined the cited references on an as-needed basis to be able to properly interpret the provisions of SCFR and the FSRG regulations under investigation.

While we did not have a preconceived notion of requirements relevance and kept an open mind about the concepts we could encounter, we were guided by two sources of information:

\begin{itemize}[leftmargin=*]
\item A literature survey on IoT food-safety monitoring \citep{Bouzembrak2019Internet}: This survey outlines five main application areas for IoT in food safety. These areas are: \emph{monitoring, sensing, communication, traceability, and data management}. Our initial review was particularly sensitive to the incidence of these concepts in the textual content being analyzed.

\item Prior research by the RE community on legal requirements: Several code sets exist for labelling and classifying legal provisions, e.g., \citep{Anton2004Requirements,Kiyavitskaya2008Automating,Bhatia2018Semantic,Sleimi2018Automated,Amaral2022AI,Torre2020Ai}. While none are specifically directed at food safety, we found the concepts of \emph{data} and \emph{data actions} (e.g., \emph{collection, retention, and transfer}) from privacy requirements \citep{Anton2004Requirements,Bhatia2018Semantic} helpful when developing our code set. Ultimately, we concluded that, in a food-safety context, there was insufficient justification for including data actions, since virtually no personal data is involved. All actions we came across in our GT data fell under either collection or retention without any clear distinction made between the two.
\end{itemize}

As discussed in Section~\ref{subsec:GT}, Straussian GT offers the flexibility to incorporate insights from existing literature, in contrast to the more purist versions of GT (e.g., Glasserian), where consulting prior literature is discouraged~\citep{Glaser1992Basics}. In our case, the flexibility to consult the literature was important, as it enabled us to be better informed about the capabilities of existing IoT technologies for food-safety monitoring and to remain sensitive to the nuances of data handling, which have been extensively studied in the RE community.

Table~\ref{tbl:CodingSummary} provides statistics on the number of initial, selected, and coded sentences from SFCR and the FSRG regulations on meat and egg. These sources combined constitute  
a total of 12,240 sentences (i.e., the initial sentences in Table~\ref{tbl:CodingSummary}). A section-by-section review of these sources resulted in the exclusion of 9,105 sentences that were not directly related to the requirements of food-safety systems. Notably, the exclusions encompass the SFCR parts that do not pertain to food products. These parts, along with brief descriptions of their content, are listed in Table~\ref{tab:SFCR_exclusion}.

\begin{table}
    \centering
    \caption{Summary of Initial, Selected, and Coded Sentences from the SFCR and FSRG Regulations.}
    \label{tbl:CodingSummary}
    \begin{tabularx}{\columnwidth}{X X X}
        \toprule
        \textbf{Initial Sentences} & \textbf{Selected Sentences} & \textbf{Coded Sentences} \\
        \midrule
        \rowcolor{lightgray}
        12,240 & 3,135 & 688 \\
        \bottomrule
        & &
    \end{tabularx}
    \noindent
    \begin{minipage}{\columnwidth}
    \raggedright
    \textbf{Initial Sentences}: Total number of sentences present in SFCR and FSRG regulations. \\
    \textbf{Selected Sentences}: Sentences that are selected for further examination based on relevance to food-safety systems. \\
    \textbf{Coded Sentences}: Sentences that are fully coded in the GT study.
    \end{minipage}
\end{table}

\begin{table}
    \centering
    \caption{SFCR Parts Excluded from Our GT Analysis on the Basis of Being Unrelated to Food-safety Systems.}
    \label{tab:SFCR_exclusion}
    \begin{tabularx}{\textwidth}{p{1.3cm} p{3.1cm} X}
        \toprule
        \textbf{Part Number} & \textbf{Part Title} & \textbf{Content}\\
        \midrule
        \rowcolor{lightgray}
        7 & Recognition of Foreign Systems & This part outlines the criteria for recognizing a foreign state's inspection system, focusing on the pertinent administrative processes.\\
        \rowcolor{white}
        8 & Ministerial Exemptions & This part addresses exemptions granted by the Minister for specific regulatory requirements, often for trade-related purposes like alleviating shortages or testing markets.  \\
        \rowcolor{lightgray}
        15 & Transitional Provisions & This part deals with the transition to SFCR from previous regulations and the associated amendments.  \\
        \rowcolor{white}
        16 & Consequential Amendments, Repeals and Coming into Force & This part lists the regulations repealed when SFCR was enacted, including the Canada Agricultural Products Act, Meat Inspection Regulations, and food requirements from the Consumer Packaging and Labelling Regulations. \\
        \bottomrule
    \end{tabularx}
\end{table}

After excluding orthogonal content, as discussed above, we were left with 3,135 sentences for further analysis (i.e., the selected sentences in Table~\ref{tbl:CodingSummary}). We began analyzing these sentences in the order of their appearance, starting with SCFR, followed by the FSRG regulation on meat, and then the FSRG regulation on egg. As we progressed through these sentences, we carefully monitored for repetitive content to ensure that our coding focused on material most likely to generate new concepts or refine existing ones. This approach allowed us to increase the diversity of the material examined while keeping the coding process manageable in terms of effort. Ultimately, 688 sentences were subjected to systematic coding, as indicated in Table~\ref{tbl:CodingSummary}.

\begin{figure}
\centering
\includegraphics[width=.95\columnwidth]{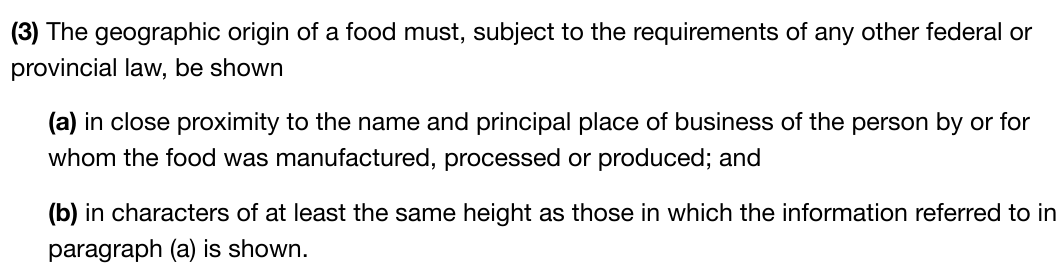}
\caption{A snippet from Part 11 of SFCR.}
\label{fig:Part11}

\vspace*{2em}

\includegraphics[width=\columnwidth]{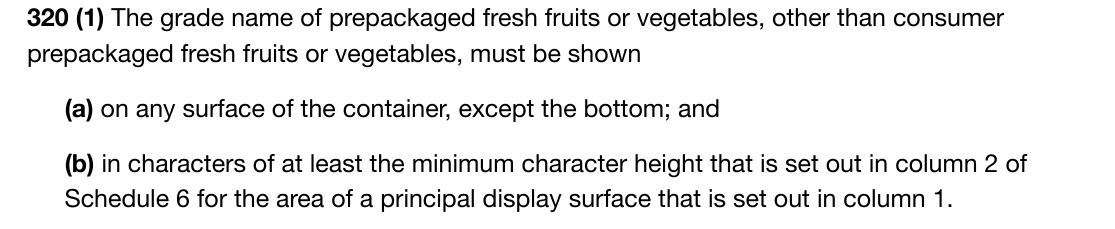}
\caption{A snippet from Part 12 of SFCR.}
\label{fig:Part12}
\end{figure}

To illustrate what we mean by repetitive content, consider the snippets in Figs. \ref{fig:Part11} and \ref{fig:Part12}. The first snippet is from Labelling (Part 11 of SFCR), while the second is from Grades and Grade Names (Part 12). Both snippets contain specific information that we classify as \emph{Label Data}. In our GT study, we opted not to code the second snippet, as the first had already been coded. This decision saved time and allowed us to focus on more novel content.

While our GT approach follows the principles of the Straussian method, we used Saldaña's practical guidelines~\citep{Saldana2015Coding} for the detailed procedures of first- and second-cycle coding to ensure consistency across multiple coding cycles. Throughout the coding process, we created a series of \emph{memos}, including notes, conceptual diagrams, and slides, to better capture and present reflections and the development of categories. Initially, first-cycle codes were generated through \emph{open coding}, aimed at identifying and categorizing key concepts related to food-safety systems, complemented by \emph{constant comparisons}. These first-cycle codes were then refined and reorganized through second-cycle \emph{axial coding}. Ultimately, the core category identified through \emph{selective coding} reflects a precise, content-driven characterization of whether a given food-safety provision is relevant to the requirements of (software-intensive) food-safety systems.
For instance, during the open coding of provisions \(P_3\) and \(P_4\) from the illustrative example in Fig.~\ref{fig:example}, we identified three preliminary concepts: data associated with food items, the collection of data about food items, and the retention of such data. Through constant comparison, the concept of food-associated data was eventually refined into \emph{Label Data}. Furthermore, recognizing that explicit data actions were unnecessary in our domain (as discussed earlier in this section), data collection and retention were merged into a single concept: \emph{Non-Label Data}. During axial coding, the concepts of \emph{Label Data} and \emph{Non-Label Data} were integrated to form the higher-order concept of \emph{Data}.

\sectopic{Coding Results.} Figure~\ref{fig:metadata} presents our content model for requirements-related provisions based on our GT study of SFCR and the FSRG regulations. This content model serves as our response to RQ1, posed in Section~\ref{sec:introduction}. The concepts in the model are the targets for automated classification, as we describe Section~\ref{sec:extraction}. The model is organized into two levels: \textit{level-1~(L1)} concepts are shaded yellow, and \textit{level-2~(L2)} concepts (i.e., sub-concepts of L1) are shaded grey. Concept definitions are provided in the glossary of Table~\ref{tbl:glossary}.  Summary statistics for concept instances identified during our GT study are provided in Table~\ref{tab:Data Representation} and discussed further in Section~\ref{subsec:dataset}.

We make four important remarks regarding the content model of Fig.~\ref{fig:metadata}:

\begin{enumerate}[leftmargin=*]
\item \textbf{Scope of Measurements.} The systems-and-software significance of measurements is in that they can in principle be made through (IoT) \emph{sensing}. Our content model covers only the measurement types observed in our GT study and is therefore not an exhaustive representation of all possible measurements. Sensing applies to various other quantities, such as pressure, luminosity, and motion, among others. 
While our GT data does not support the inclusion of additional measurements, it does not rule them out either. A broader GT study may find new measurements relevant for food-safety management. Should other measurement types be found pertinent, we do not foresee saturation problems in our content model, noting that
the number of fundamental physical quantities that can be measured is limited.

\begin{figure}[!t]
\centering
\includegraphics[width=.7\columnwidth]{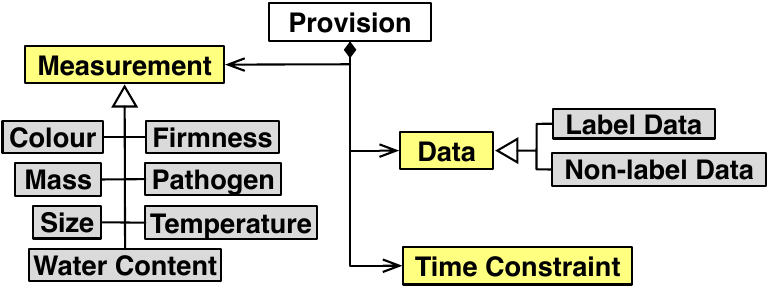}
\caption{Content model for requirements-related provisions in food-safety regulations.}
\label{fig:metadata}
\end{figure}

\begin{table}[!t]
\vspace*{1em}
\caption{Glossary for the Content Model of Fig.~\ref{fig:metadata}.}
\label{tbl:glossary}
\fontsize{8.3}{8.3}\selectfont
\renewcommand{\arraystretch}{1.65}
\begin{tabularx}{\columnwidth}{p{2.6cm} X}
  \toprule
  \textbf{Concept} & \textbf{Definition}\\
  \midrule
  \rowcolor{lightgray}
  \vspace*{.5em}
  \emph{Data}\linebreak (+sub-concepts)  & 
   \emph{Data:} any information used to convey knowledge, provide assurance or perform analysis;
   \emph{Label Data:} information that a food-product package or container must bear; 
   \emph{Non-label Data:} any food-safety-relevant data other than label data that needs to be collected and/or retained for inclusion in documents such as certificates, reports, guarantees and letters.\\
\rowcolor{white}
  \emph{Measurement}\linebreak (+sub-concepts) & \emph{Measurement:} Association of numbers with physical quantities; \emph{Colour:} (self-evident);
\emph{Firmness:} degree of resistance to deformation;
\emph{Mass:} amount of substance by weight or volume;
\emph{Pathogen:} a microorganism that causes disease;
\emph{Size:} dimension (e.g., length or thickness) or surface area;
\emph{Temperature:} (self-evident);
\emph{Water Content:} humidity or moisture.\\ 
\rowcolor{lightgray}
\emph{Time Constraint} & A temporal restriction, which in our context, is expressed using intervals, deadlines or periodicity. \\
  \bottomrule
\end{tabularx}
\end{table}

\item \textbf{Implicit Concepts.} Although traceability, supply chain and communication are key concepts for food-safety management \citep{Bouzembrak2019Internet}, our GT study did not result in explicit codes for these concepts. Given the technology-neutral nature of SFCR, this finding is not surprising for supply chain and communication, which are technology-laden. The absence of express traceability concepts nevertheless led to further investigation into the underlying reasons. SFCR has a dedicated section on traceability (Part 5). Our follow-up examination of this section confirmed that SFCR establishes traceability requirements primarily through \emph{indirect} means. Specifically, traceability is achieved through the use of information provided on food-product labels (such as lot codes or unique identifiers) and through documentation submitted by distributors, transporters, importers, exporters and retailers. Both types of information are represented in our content model through the concepts of \emph{Label Data} and \emph{Non-label Data}.

\item\textbf{Concept Refinement and Consolidation.} The concepts in the model of Fig.~\ref{fig:metadata} emerged only after several rounds of refinement. Initially, our set of open codes included the following 18 concepts: \emph{Data Collection}, \emph{Data Retention}, \emph{Data Definition--Food-attached}, \emph{Temperature}, \emph{Dimension Measure}, \emph{Weight/Capacity}, \emph{Water Level}, \emph{Moisture}, \emph{Firmness}, \emph{Bacteria}, \emph{Tests}, \emph{Fat Thickness}, \emph{Colour}, \emph{Volume}, \emph{Deadline/Interval}, \emph{Periodicity}, \emph{Font}, and \emph{Format}. Through iterative analysis, we identified conceptual overlaps, leading to the merging, renaming, or discarding of several initial codes. For example, \emph{Volume} and \emph{Weight/Capacity} were merged into \emph{Mass}, while \emph{Fat Thickness} was discarded, as it was subsumed under the broader concept of \emph{Size}. In total, eight concepts were consolidated into four, and five concepts were discarded. The remaining codes were refined to align more closely with the main regulatory concepts identified: (1)~\emph{Dimension Measure} was renamed \emph{Size} to simplify and generalize the concept of physical dimensions for food products, and (2)~\emph{Data Definition--Food-attached} was renamed \emph{Label Data} to more accurately reflect its direct application to food products.

\item
\textbf{Context-driven Codes for Operationalization.} Food-safety regulations focus on \emph{what} must be done, not \emph{how}. For instance, the regulations might require the measurement of \emph{Size} without specifying the method, leaving it open to, say,  use calipers for fat thickness versus a thickness gauge for sliced products like cheese and meat. Our GT study reflects this by identifying concepts applicable across various food products. However, different products have unique characteristics, so to generate ``lower-level'' codes that can help address the ``how'' for operationalization, one must consider \emph{tuple}-combinations of contextual information about a food product and a high-level code from our content model of Fig.~\ref{fig:metadata}. For instance, a tuple such as (``citrus fruit'', \emph{Size Measurement}) provides the necessary context for determining the appropriate method of measurement. We observe that no significant human effort is required to identify the relevant food product in a specific provision: When a provision targets a particular product, this is clearly indicated by the structure of the regulation. Typically, section titles help analysts easily distinguish between provisions for categories like meat, egg, and fruits. Thus, lower-level codes can be inferred from the \emph{Cartesian product} of food products and the high-level codes identified in our GT study. Explicitly enumerating the Cartesian product of all potential combinations (food products and high-level codes) would not provide additional insight to our study, as the regulations themselves avoid irrelevant combinations, e.g., (``milk'' and \emph{Firmness Measurement}).  Ruling out what is illogical or unhelpful (from a regulatory standpoint) is easily achievable once a given snippet of regulatory text for a specific food product has been processed. For instance, after automatically analyzing and labelling the section concerned with dairy products, the user can pose a query such as ``Are there any statements in this content that have been annotated by the concept of \emph{firmness}?
\end{enumerate}

\section{Automated Classification of Food-safety Provisions} \label{sec:extraction}
In this section, we present a pipeline for  automatic classification of regulatory food-safety provisions based on the concepts identified in our GT study of Section~\ref{fig:metadata}. Figure~\ref{fig:approach} presents an overview of the pipeline, which takes as input the textual content of the regulations being examined and produces as output labels for each provision in the input.

The pipeline treats each sentence as one provision. Stated otherwise, our \emph{unit of analysis} is a sentence. This decision aligns with previous research on automated processing of legal texts~\citep{Amaral2022AI}, which found that sub-sentence localization offers limited additional value, while larger units like paragraphs and sections may lack precision. 
Although \citet{Amaral2022AI} focus on data-protection regulations, we observe that food-safety regulations share key characteristics with other legal frameworks, including data protection. The similarities notably include formal legal language, defined roles, distinctions between permissions and obligations, and accountability criteria. In food safety, actors such as food producers, manufacturers, and regulatory bodies (e.g., the Canadian Food Inspection Agency) ensure product safety, proper labelling, and hygiene. Likewise, data protection frameworks assign roles to data controllers and processors, overseen by data protection authorities, as in the General Data Protection Regulation (GDPR). Both regulatory frameworks differentiate between permissible actions, like using approved food additives or processing personal data lawfully, and obligatory actions, such as maintaining hygiene or protecting personal data. Accountability mechanisms, such as audits and inspections, are central to both. These commonalities support using sentences as natural units of analysis for classifying food-safety regulations. 

Our pipeline has four steps, as we describe next:
  
\begin{figure}
\centering    \includegraphics[width=.8\linewidth]{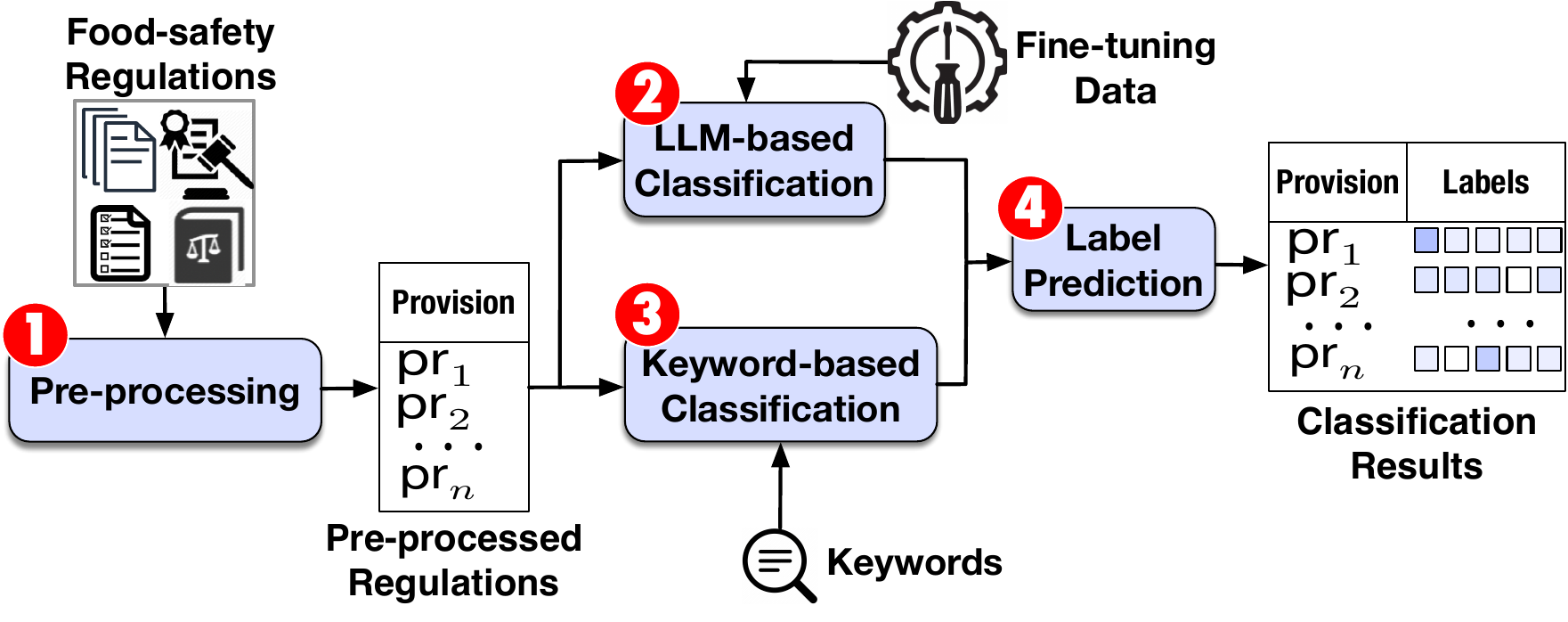}
\caption{Overview of automated classification pipeline.}\label{fig:approach}
\end{figure}

\subsection{Pre-processing} In this step, we split the input into sentences. The term ``sentence'' does not necessarily refer to a grammatical sentence, but rather to a text segment that has been identified as a sentence by the sentence splitter module in the NLP pipeline. Legal documents frequently use lists. To ensure preservation of context during sentence splitting, we follow the approach of \citet{Sleimi2018Automated} and add the header in each list, known as a list-item prefix, to each itemized/enumerated item.

\subsection{LLM-based Classification}
Using an LLM customized using labelled data from our GT study (Section~\ref{sec:metadata}), we classify provisions with occurrences of concepts from the model of Fig.~\ref{fig:metadata}. 
In our evaluation (Section~\ref{sec:evaluation}), we consider four families of models -- BERT, GPT, Llama, and Mixtral -- as alternatives. For the generative LLMs in our evaluation, the pre-processed regulations from Step~1 are embedded into prompts, as explained in Section~\ref{subsec:evaluationprocedure}.

\subsection{Keyword-based Classification}\label{subsec:keyword-based}
Our GT study did not yield a sufficiently large number of examples for \emph{Colour}, \emph{Firmness}, \emph{Pathogen} and \emph{Water Content}. As previous studies on automated classification of legal texts point out \citep{Sleimi2018Automated,Torre2020Ai,Amaral2022AI}, some concepts may turn out to be too scarce for learning-based techniques, thereby warranting the use of keywords for classification. In the third step of our approach, we perform a keyword lookup for the above four concepts, for which we found too few ($\leq$10) instances in our GT study. 
The keywords associated with each concept were derived during our GT study and are listed in our online material \citep{Data}. If a provision $P$ contains one or more of the keywords associated with a concept $M$, we label $P$ with $M$. For instance, if a provision contains the keywords ``E. coli'' or ``Salmonella Enterica'', we label it with \emph{Pathogen}. We note that this keyword-based strategy is reserved exclusively for sparse concepts due to the substantial effort needed to develop comprehensive keyword lists as well as the risk of overfitting.

\subsection{Label Prediction} This step combines the labels computed by the LLM-based classifiers (Step~2) and those computed by the keyword-based classifiers (Step~3) to produce the final label recommendations for each provision. 

\section{Empirical Evaluation}\label{sec:evaluation}
In this section, we instantiate the classification pipeline of Fig.~\ref{fig:approach} using various LLMs and customization strategies, and then evaluate the accuracy of each resulting configuration.

\subsection{Implementation}\label{subsec:implementation}
Our classification pipeline is implemented in Python. The sentence splitter is implemented using SpaCy 3.4.4. 
The BERT models we experiment with are: \emph{albert-base-v1}, \emph{bert-large-cased}, \emph{bert-base-cased}, 
and \emph{roberta-base}, all from Hugging Face \citep{HuggingFace} and operated in PyTorch 1.13.1.
To optimize the fine-tuning hyperparameters of BERT, we use Optuna 3.1.0 \citep{Akiba2019Optuna}.
The GPT models we experiment with are \emph{GPT-3.5-turbo} and \emph{GPT-4o}, the Llama model is \emph{Llama-3-8B-Instruct}, and the Mixtral model is \emph{Mixtral-8x7B-Instruct-v0.1}. For fine-tuning GPT, we use the OpenAI API through the OpenAI Python package. For fine-tuning Llama and Mixtral, we use Transformers 4.40.1, BitsAndBytes 0.42.0 for 4-bit model quantization, PEFT 0.12.0 for LoRA-based fine-tuning, and TRL 0.9.6 for supervised fine-tuning.

We further implement two baseline approaches for comparison: one utilizing the (Bi)LSTM architecture \citep{Graves2005Framewise} and the other using automatic keyword extraction. 
The BiLSTM baseline is implemented in PyTorch 1.13.1, with the required NLP pipeline and word embeddings built using Keras 2.11.0 and GloVe \citep{Pennington2014Glove}. We use the same approach for hyperparameter optimization of the BiLSTM baseline as we did for BERT.
Our keyword-based baseline employs the Python-based implementation of \citet{Arora2016Automated}'s keyword extraction approach, available as part of the WikiDoMiner tool \citep{Ezzini2022WikiDoMiner}. This baseline trains a Random Forest (RF) classifier to learn how to associate certain keyword combinations with a specific label. Once trained, the classifier can be used to predict the label of new sentences. This baseline further uses scikit-learn 1.2.0 to implement both its Count Vectorizer and RF classifier. We note that this baseline is not comparable to Step~3 of our classification pipeline (Section~\ref{sec:extraction}), which requires manually defined keywords and is reserved for sparse concepts only. In addition, we use scikit-learn 1.2.0 for dataset splitting and metrics calculation throughout our evaluation.
As a technical remark, it is worth mentioning that for BERT and our two baselines, we have structured our requirements-relevance classification problem -- a multi-label classification task -- as a series of binary classification tasks, with each binary classifier predicting the presence or absence of one specific concept within provisions.
All our implementation artifacts are available online \citep{Code}.

\subsection{Data Collection Procedure}\label{subsec:datacol}

\begin{sloppypar}Our evaluation dataset consists of two distinct components: a fine-tuning/training set, denoted $F$, and a test set, denoted $T$.
Set $F$ comes from our GT study of Section~\ref{sec:metadata}.
For further information on our GT study that led to the development of 
$F$, we have made available our initial labelling document, which includes guidelines, colour-coded sheets, raised questions, notes, and relevant links, as part of our online material \citep{Data}. Below, we describe the procedure for building $T$. 
As explained in Section 3, we excluded from our qualitative study all FSRG regulations except those on meat and egg products.
To construct $T$, we include the following FSRG regulations: dairy products, fish, fruits and vegetables, honey, manufactured products, and maple. These are all the FSRG regulations, excluding those on meat and egg, which were part of our GT study. In addition, we include selected food-safety regulations from the US Food and Drug Administration (FDA).
The motivation to augment our test set with FDA regulations is to assess the generalizability of our pipeline over food-safety regulations in a jurisdiction different from that covered by the documents examined in our GT study. This augmentation is also natural and useful, considering that, like Canada, the US ranks highly in ``Quality and Safety'' according to GFSI \citep{GFSI}, making it a valuable source for evaluating the thoroughness of our classification experiments.\end{sloppypar}

For the FDA portion, we consider three sources under the food category: \emph{packaging} \citep{Packaging2017}, \emph{labeling} \citep{Labeling2022} and \emph{guidance for industry}~\citep{Safety2022,FCS2019}. 
To create $T$, we randomly select 400 provisions (sentences). Of these, 350 are from the FSRG regulations excluded from our GT study, and the remaining 50 from the FDA sources cited above.
To create an unbiased ground truth for testing, we collaborated with two third-party annotators (non-authors). Both are graduate students in Computer Science. They possess excellent English language skills, have 2+ years of prior industry experience, and have been exposed to requirements engineering and qualitative coding in their graduate coursework. \hbox{The annotators were paid for their work.}

We conducted a two-hour training session for the annotators before they started their work. The training covered food-monitoring systems, food-safety management and the content model of Fig.~\ref{fig:metadata}. To label provisions according to this content model, the annotators were instructed to apply hypothesis coding -- a qualitative coding approach in which pre-existing codes guide the analysis~\citep{Saldana2015Coding}. In this case, the pre-existing codes are those from the content model of Fig.~\ref{fig:metadata}. The annotators received a protocol document to use as a reference, explaining concepts of interest with examples; this protocol document is available in our online material~\citep{Data}. While the annotators were given the flexibility to extend our codes if needed to incorporate new insights, no new codes emerged from their work.
 
We divided the 400 provisions in $T$ equally between the two annotators, including a 10\% overlap to assess annotation reliability. 
That is, each annotator examined 210 provisions (190 distinct and 20 shared). We use Cohen's kappa ($\kappa$) for measuring inter-rater agreement.
For a concept $M$, a provision $P$ counts as an agreement if both annotators make the same decision as to whether $M$ applies to $P$. \hbox{Divergent decisions count as disagreements.}

\subsection{Evaluation Metrics}\label{subsec:metricS}
We evaluate accuracy using \emph{Precision} and \emph{Recall}, with their standard definitions. For every classification label $M$, the definitions of True Positive (TP), False Positive (FP), True Negative (TN), and False Negative (FN) are as follows: A \emph{TP} is a prediction of \textsf{true} for a provision that has $M$ in its label set (as per the ground truth); a \emph{FP} is a prediction of \textsf{true} for a provision that does not have $M$ in its label set; a \emph{TN} is a prediction of \textsf{false} for a provision that does not have $M$ in its label set; and a \emph{FN} is a prediction of \textsf{false} for a provision that has $M$ in its label set.

Precision and Recall are often reported alongside the \emph{$F$ measure} (or \hbox{\emph{$F$ score}}) -- the harmonic mean of Precision and Recall -- which provides a unified metric for classification accuracy. For clarity, we discuss accuracy directly in terms of Precision and Recall. F-measure results are available in our online material~\citep{Data}.

We use the non-parametric Wilcoxon Rank-Sum Test~\citep{Capon1991Elementary}, also known as Mann-Whitney U Test, to compare the accuracy of different algorithms, specifically different classification pipelines in our context. Through this test, we assess whether there is a significant difference in the distribution of accuracy metrics between two algorithms, without assuming any particular underlying data distribution. We note that, in our evaluation, we always compare alternative algorithms by measuring their accuracy on a single test set. The reason we obtain a distribution rather than a single-point accuracy value is the inherent random variation caused by the non-deterministic nature of training, fine-tuning, and (potentially) inference in deep learning, which typically leads to a range of accuracy values. For example, even when BERT or GPT is fine-tuned multiple times under the same conditions, randomness in the fine-tuning process can result in different model parameters. This can in turn cause the models to yield different accuracy results when tested on the same dataset. In generative LLMs like GPT, this effect is intensified by hyperparameters such as temperature, which introduce randomness during output generation.

The null hypothesis ($H_0$) for our statistical tests states that there is no difference in the distribution of the accuracy metric of interest -- either \emph{Precision} or \emph{Recall} -- between the two algorithms being compared. The goal of these tests is to assess whether the observed differences are statistically significant or merely due to random variation~\citep{arcuri2011practical}.

To assess the practical significance of the difference between two algorithms, we use Vargha-Delaney's non-parametric effect-size statistic, denoted as $\hat{A}_{12}$~\citep{Vargha2000Critique}. Given a pair of algorithms, the value of $\hat{A}_{12}$ represents the probability that a randomly selected observation from one algorithm will outperform a randomly selected observation from the other. We categorize effect sizes into four levels, based on the widely used threshold-based guidelines from \citet{Vargha2000Critique}: \textit{L (Large)}: A strong difference between the algorithms, indicating a clear performance advantage; \textit{M (Medium)}: A moderate difference, suggesting some advantage, though less pronounced; \textit{S (Small)}: A small difference, indicating limited practical difference; and \textit{N (Negligible)}: Almost no practical difference between the algorithms.

\subsection{Analysis Procedure} \label{subsec:evaluationprocedure}
\sectopic{EXPI.} This experiment answers RQ2 as posed in Section~\ref{sec:introduction}. The experiment consists of two stages: (1) customizing an LLM using the $F$ portion of the dataset, as defined in Section~\ref{subsec:datacol} and (2) applying the (customized) LLM to the $T$ portion, as defined in the same section. 
The technical details of these two stages differ depending on the chosen LLM and whether fine-tuning or few-shot learning is used for customization. We describe the stages separately for BERT, GPT, Llama and Mixtral.

Given the non-determinism in training deep learning models, caused by factors such as initialization and regularization, it is important to consider random variation during both the initial training phase and any subsequent fine-tuning~\citep{Pham2020Problems,Chen2022Towards}. In generative LLMs like GPT, beyond the stochastic elements of training, inference also involves sampling from a probability distribution. As mentioned in Section~\ref{subsec:metricS}, hyperparameters such as temperature, which determine how deterministic or creative the output of an LLM is, introduce additional variability.
To account for random variation, we conduct 20 independent experiments for each concept $M$ in our content model (Fig.~\ref{fig:metadata}) and present the results as boxplots\footnote{As we discuss in Section~\ref{subsec:RQ2}, challenges with Llama and Mixtral in getting them to follow instructions prevented systematic experimentation with these two LLM families. Therefore, we provide detailed results only for BERT and GPT.}.

\begin{sloppypar}\underline{\emph{BERT.}} We experiment with four BERT variants: BERT base, BERT large, ALBERT and RoBERTa. To fine-tune each BERT model effectively, we optimize three hyperparameters: learning rate, number of epochs, and choice of optimizer. The ranges explored for learning rate and the number of epochs are $[2\mathrm{e}{-5}, 2\mathrm{e}{-4}]$ and $[10, 35]$, respectively, in accordance with best practices \citep{Devlin2018Bert}. For the optimizer, we consider two alternatives: SGD \citep{Leon1998Online} and AdamW \citep{Loshchilov2017Fixing}. 
As for the batch size and maximum sequence size hyperparameters, we respectively use 16 and 512 -- the default values suggested by \citet{Devlin2018Bert}.
The optimal hyperparameters for each concept of our model are provided online \citep{EvaluationResults}.
With the BERT hyperparameters optimized and the most accurate fine-tuned model built for each concept in the first stage, the second stage applies the resulting BERT models to the (third-party-annotated) test set, $T$.\end{sloppypar}

\underline{\emph{GPT.}} We experiment with two variants of GPT: GPT-3.5-turbo and GPT-4o. To adapt these models to our task-specific data, we consider fine-tuning. A natural question that may arise here though is whether fine-tuning powerful LLMs such as GPT is worthwhile, considering that these models already have extensive pre-trained knowledge and generalization capabilities. To be able to address this question, in our evaluation of GPT-4o, we further explore \emph{few-shot learning} as an alternative to fine-tuning.

\begin{figure}[ht]
\includegraphics[width=\linewidth]{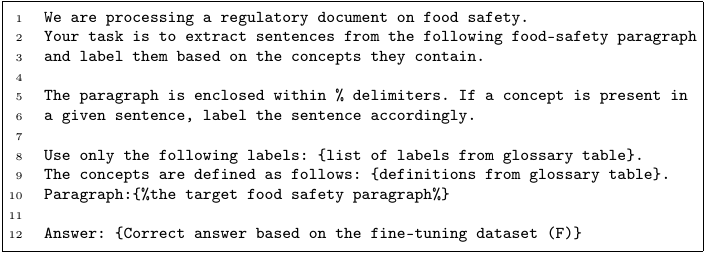} 
\caption{Prompt used for fine-tuning. Once the model is fine-tuned, to have it predict labels for a given paragraph, line 12 needs to change to ``Answer.'' (in the imperative form).}
\label{fig:fine-tuned prompt}
\end{figure}

Fine-tuning is done over three epochs, with the number selected automatically by GPT based on our dataset size. The fine-tuning data, $F$, is fed to GPT in a conversational chat format, where examples are structured as messages containing roles for the system, user, and assistant \citep{OpenAiFinetuning}. The system role provides instructions that the model should follow; the user role presents input that the model should respond to; and, the assistant role represents the model's response to the user's input. In this setup, the user role presents input paragraphs, and the assistant role outputs the labelled sentences. After fine-tuning, the fine-tuned models are evaluated against $T$.
In our context, the system role instructs the model to label the sentences based on the identified concepts.  The user role and the assistant role present input paragraphs and the labelled sentences, respectively.
Our prompt for fine-tuning is shown in Fig.~\ref{fig:fine-tuned prompt}.

\begin{figure}[ht]
\includegraphics[width=\linewidth]{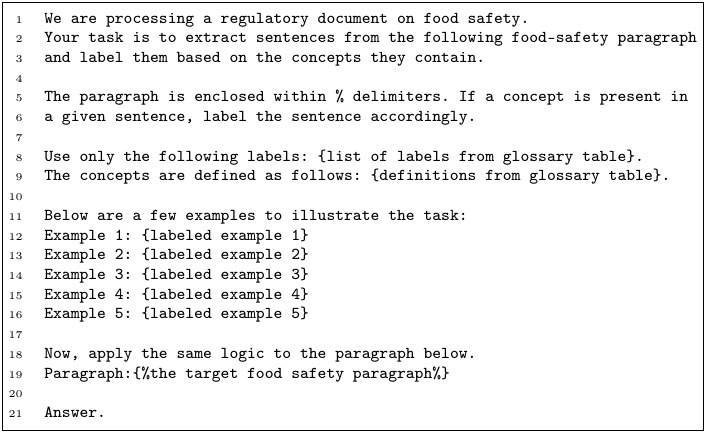} 
\caption{Prompt used for few-shot learning.}
\label{fig:few-shot-prompt}
\vspace*{-1em}
\end{figure}

We set GPT's temperature hyperparameter to a low value, 0.2, following OpenAI's guidelines for situations where one wants to minimize variation in responses and thus make GPT outputs more deterministic~\citep{Temperature}. While increased determinism is generally advantageous for classification tasks, absolute determinism is often not ideal, and in fact not technically feasible in GPT either, as per OpenAI's documentation~\citep{Temperature}. To this end, we further note that our classification task is text classification. Textual descriptions, especially laws and regulations, often contain ambiguities and context-dependent nuances. Attempting to eliminate non-determinism could therefore prevent the model from accounting for different plausible interpretations.

For few-shot learning, GPT-4o is provided with a small set of examples to help it understand the classification task. Specifically, we include in our prompt five representative examples taken from the fine-tuning data, $F$. Our few-shot prompt (excluding the five examples) is shown in Fig.~\ref{fig:few-shot-prompt}. We note that explicit output formatting instructions are not included in either the fine-tuning or few-shot prompts. This is because the examples provided in both contexts capture both the classification task \emph{and} the desired output format, allowing us to keep output formatting implicit.

\underline{\emph{Llama and Mixtral.}} We experiment with one variant of Llama, specifically Llama-3-8B-Instruct~\citep{Llama}, and one variant of Mixtral, specifically Mixtral-8x7B-Instruct-v0.1~\citep{Mixtral}. These variants were chosen because they are optimized for instruction-following, and they were the largest models that our institutional computational resources could accommodate. We fine-tune these models over 25 epochs using Parameter-Efficient Fine-Tuning (PEFT) with Low-Rank Adaptation (LoRA). The hyperparameters \texttt{r} and \texttt{lora\_alpha} are set to their default values, 16 and 32, respectively, as provided by Hugging Face's PEFT framework~\citep{PEFT}. In both models, the temperature hyperparameter is set to 0.2 (the same value used for the GPT models). The fine-tuning set, \(F\), is fed to the models in a conversational chat format, similar to GPT fine-tuning (Fig.~\ref{fig:fine-tuned prompt}), but adapted to the specific prompt structures of these models~\citep{Llama,Mixtral}.

For Llama, the system, user, and assistant roles are structured similarly to GPT, but with model-specific tokens to distinguish them in the input prompt. In contrast, Mixtral uses a simpler format that does not explicitly separate system and user roles. Instead, the prompt consists of an instruction followed by the model's response. Detailed prompts for both Llama and Mixtral can be found in our online material~\citep{Code}.

\vspace*{.2em}\sectopic{EXPII.} This experiment answers RQ3, as posed in Section~\ref{sec:introduction}. We compare our LLM-based classification models against two non-Transformer (and simpler) models, namely \emph{BiLSTM} and \emph{automatic keyword extraction}, as we outline below.

The input to the BiLSTM baseline is a sequence obtained by tokenizing the food-safety provisions to classify. The tokens are represented using GloVe word embeddings \citep{Pennington2014Glove}. We optimize the hyperparameters of the BiLSTM model and train the model (on $F$) in a manner similar to BERT, as explained in EXPI. The implementation of our BiLSTM baseline, the set of hyperparameters for this baseline, and the optimal values of the hyperparameters obtained for each concept in our content model are provided online \citep{EvaluationResults}.

Our keyword-based baseline is inspired by the well-known bag-of-words classifier \citep{Joachims2005Text,Falkner2019Identifying}. Instead of using tokens as features, however, this baseline uses keyphrases to represent the content. The baseline first uses the approach of \citet{Arora2016Automated} to extract domain-specific keyphrases from the $F$ portion of our dataset. It then builds a Random Forest (RF) classifier (over $F$) using the extracted keyphrases as features.

Like the models in EXPI, our baselines are affected by random variation. In the case of BiLSTM, the source of randomness is the same as that for BERT models. For Keyword Search, the randomness is caused by the ``random state'' variable in the RF algorithm \citep{RandomForestRandomness}. To account for randomness, we employ the same methodology as in EXPI, running each baseline 20 times on the test set $T$ and reporting boxplot statistics.

\subsection{Results}\label{sec:result}
\subsubsection{Dataset}\label{subsec:dataset}

We recall from Section~\ref{subsec:datacol} that our dataset is comprised of a fine-tuning/training set, $F$, and a test set, $T$. Set $F$ resulted from our GT study (Section~\ref{sec:metadata}) and set $T$ was developed by third-parties. We observed only one discrepancy in the level-2 labels that the annotators provided for the overlapping part of their tasks. One annotator identified \emph{Label Data} in one of the provisions whereas the other annotator did not. This resulted in a Cohen's $\kappa$ of $\approx$0.90 (almost perfect agreement) for \emph{Label Data}. All other $\kappa$ values are 1 when at least one concept instance was present. 
Table~\ref{tab:Data Representation} provides summary statistics for the incidence of different concepts in $F$ and $T$. For $T$, we further show the breakdown between labels over the FSRG and FDA regulations.

\begin{table}
\centering
\caption{Distribution of Labels in our Dataset.}
\label{tab:Data Representation}
  \begin{tabular} {|p{0.22\textwidth}|p{0.125\textwidth}|p{0.25\textwidth}|p{0.13\textwidth}|}
        \hline

\multirow{2}{*}{\bf Label\,(Level)} & \multirow{2}{0pt}{\bf  GT Study~($F$)} & 
\multicolumn{2}{c|}{\bf Third-party Annotations ($T$)}\\\cline{3-4}
& & \bf Canada\,(FSRG\,Reg.) & \bf USA\,(FDA)\\
\hline
\rowcolor{lightgray}
\emph{Colour}\,(L2)            & 10        &  10  &  0              
\\ \hline
\emph{Data}\,(L1)           & 178        &  69  &   11            
\\ \hline
\rowcolor{lightgray}
\emph{Firmness}\,(L2)            &5       &   2 &0
\\  \hline
\emph{Label Data}\,(L2)          & 121        &   53 &  10           
\\ \hline
\rowcolor{lightgray}
\emph{Mass}\,(L2)           & 94        &  65  &   32              
\\ \hline
\emph{Measurement}\,(L1)            & 165        &  109  &   33          
\\ \hline
\rowcolor{lightgray}
\emph{Non-label\,Data}\,(L2)\hspace*{-1em} & 57        &  18  &  1   
\\ \hline
\emph{Pathogen}\,(L2)          & 10        &  4  &  0           
\\ \hline
\rowcolor{lightgray}
\emph{Size}\,(L2)           & 32        &  36  &    0           
\\ \hline
\emph{Temperature}\,(L2)      & 31        & 13   &  1             
\\ \hline
\rowcolor{lightgray}
\emph{Time\,Constraint}\,(L1)   & 44        &   9 &   8  
\\ \hline  
\emph{Water\,Content}\,(L2)          & 3        &  4  & 0
\\ \hline
\rowcolor{SkyBlue}  
\emph{Overall}            &369        &   184 & 40
\\ \hline
\end{tabular}
\end{table}

Of the 688 provisions in $F$, 369 ($\approx$54\%) have some label attached to them; and, of the 400 provisions in $T$, 224 (56\%) have some label attached. The \emph{Overall} label in Table~\ref{tab:Data Representation} represents a higher-level concept implied by the presence of more specific concepts in the content model of Fig.~\ref{fig:metadata}. More precisely, a given provision receives the \emph{Overall} label if and only if it has been marked with at least one concept from the content model. Essentially, \emph{Overall} distinguishes between provisions that have implications for software (i.e., those annotated with some concept from the model of Fig.~\ref{fig:metadata}) and those that are not relevant to software. The ability to make this distinction directly supports Use~Case~1, as discussed in Section~\ref{sec:introduction} (Target Users and Practical Applications). For classification, we treat \emph{Overall} as an independent label.

Four concepts, \emph{Colour}, \emph{Firmness}, \emph{Pathogen} and \emph{Water Content} are scarce in our GT study. As noted in Section~\ref{subsec:keyword-based}, we consider using keywords identified in our GT study to classify these four concepts.

\subsubsection{RQ2: How accurately can LLMs classify requirements-related provisions in food-safety regulations?}\label{subsec:RQ2}

\sectopic{Identifying the most accurate BERT variant.} Table~\ref{tab:BERT_variants} presents the average (Avg) and standard deviation (SD) for accuracy metrics, calculated over 20 runs of the four BERT variants under consideration, on the test set, $T$. Hyperparameter optimization and fine-tuning for all variants followed the procedure in Section~\ref{subsec:evaluationprocedure}. Among the variants, with one exception, BERT base yields the best accuracy metrics for the \emph{Overall} concept.
The exception is RoBERTa's \emph{Recall}, which is 1\% better than that of BERT base. However, this gain in \emph{Recall} is small and without statistical significance. In general, if one considers the  metrics of Table~\ref{tab:BERT_variants} for the \emph{Overall} concept only, there is little difference to note between the alternatives in Table~\ref{tab:BERT_variants}. A closer examination of the behaviour of the variants on individual labels nonetheless reveals that BERT base has the least amount of random variation. We select BERT base as the best-performing variant for further comparison with generative LLMs later in this section. We refer the reader to our online material~\citep{EvaluationResults} for detailed results on the other three BERT variants.

\sectopic{Challenges with Llama and Mixtral.}
We spent significant time and computational resources on fine-tuning and testing the Llama and Mixtral variants considered (Llama-3-8B-Instruct~\citep{Llama} and Mixtral-8x7B-Instruct-v0.1~\citep{Mixtral}). However, we were unable to achieve satisfactory results.
These models frequently introduced unnecessary justifications, repeated their responses, and paraphrased or generated content not found in the input. These issues made systematic evaluation impossible because the outputs were unpredictable. We believe these challenges stem from the complexity of our classification task, which requires: (1) reading a text passage and extracting sentences based on knowledge learned during fine-tuning on the training set, (2) understanding our conceptual model's customized definitions, and (3) accurately assigning correct hierarchical labels.

As a result, we exclude Llama and Mixtral from RQ2 and do not report accuracy statistics for them. Our analysis of generative LLMs therefore focuses on the GPT family and compares it with BERT base, the best-performing model within the (non-generative) BERT family.

\begin{table}
\vspace*{1em}
\caption{Accuracy of BERT variants for the \emph{Overall} concept}
\label{tab:BERT_variants}
\centering
\includegraphics[width=0.35\linewidth]{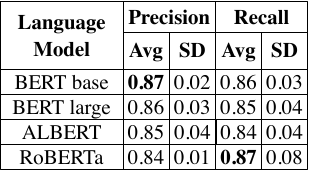}
\end{table}

\begin{figure*}
\centering
\hspace*{-4em}\includegraphics[width=1.35\linewidth]{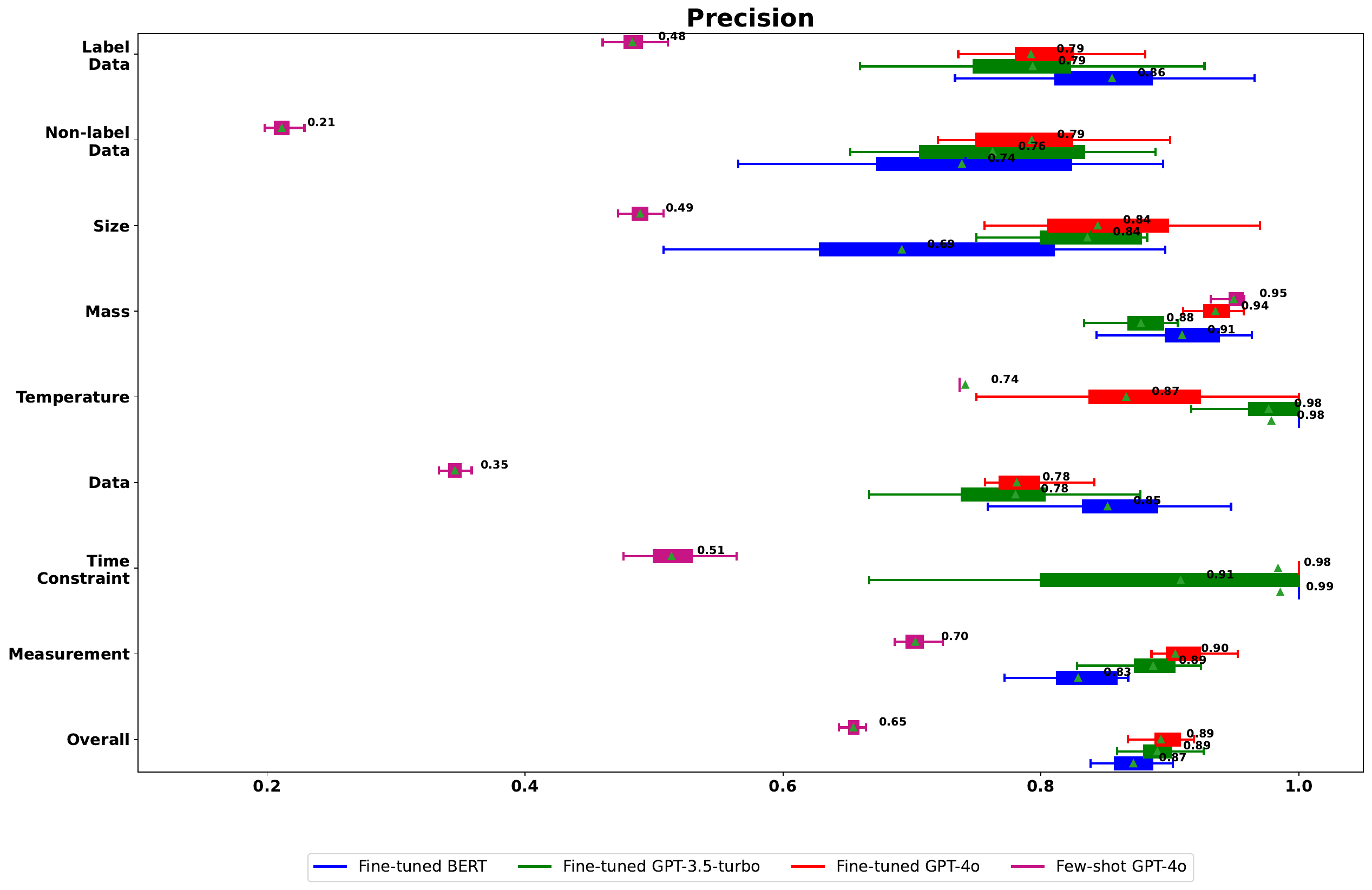}
\caption{\emph{Precision} for fine-tuned BERT base, fine-tuned GPT-3.5-turbo, fine-tuned GPT-4o, and GPT-4o with few-shot learning (the $x$-axis represents the \emph{Precision} values).}
\label{fig:Precision}

\centering\hspace*{-4em}\includegraphics[width=1.35\linewidth]{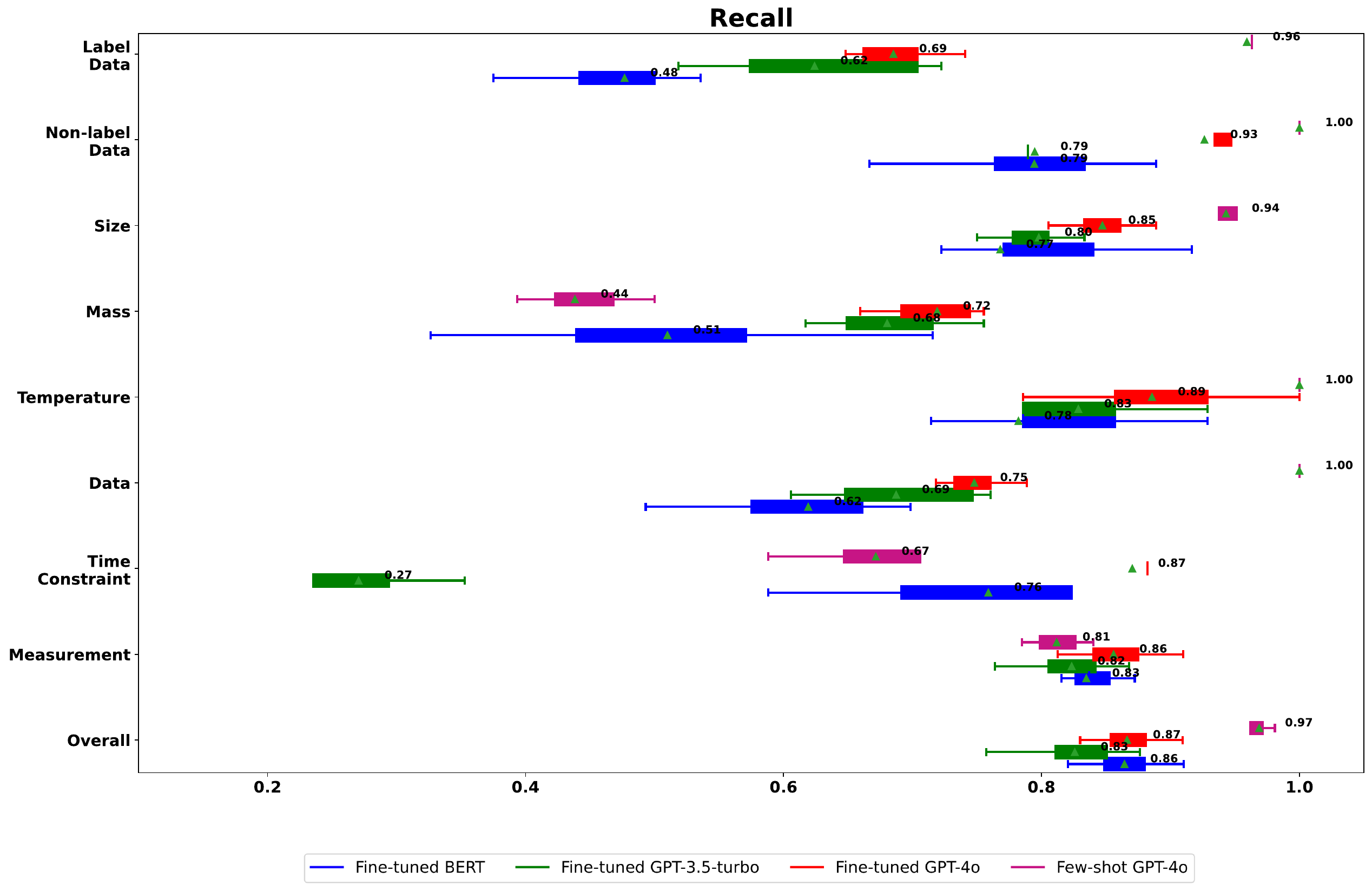}
\caption{\emph{Recall} for fine-tuned BERT base, fine-tuned GPT-3.5-turbo, fine-tuned GPT-4o, and GPT-4o with few-shot learning (the $x$-axis represents the \emph{Recall} values).}
\label{fig:Recall}
\end{figure*}

\begin{table}
\vspace*{.5em}
\caption{Statistical Comparisons for RQ2 and RQ3.}
\label{tab:Statistical Significance Testing}
\centering
\hspace*{-2em}\includegraphics[width=1.05\linewidth] {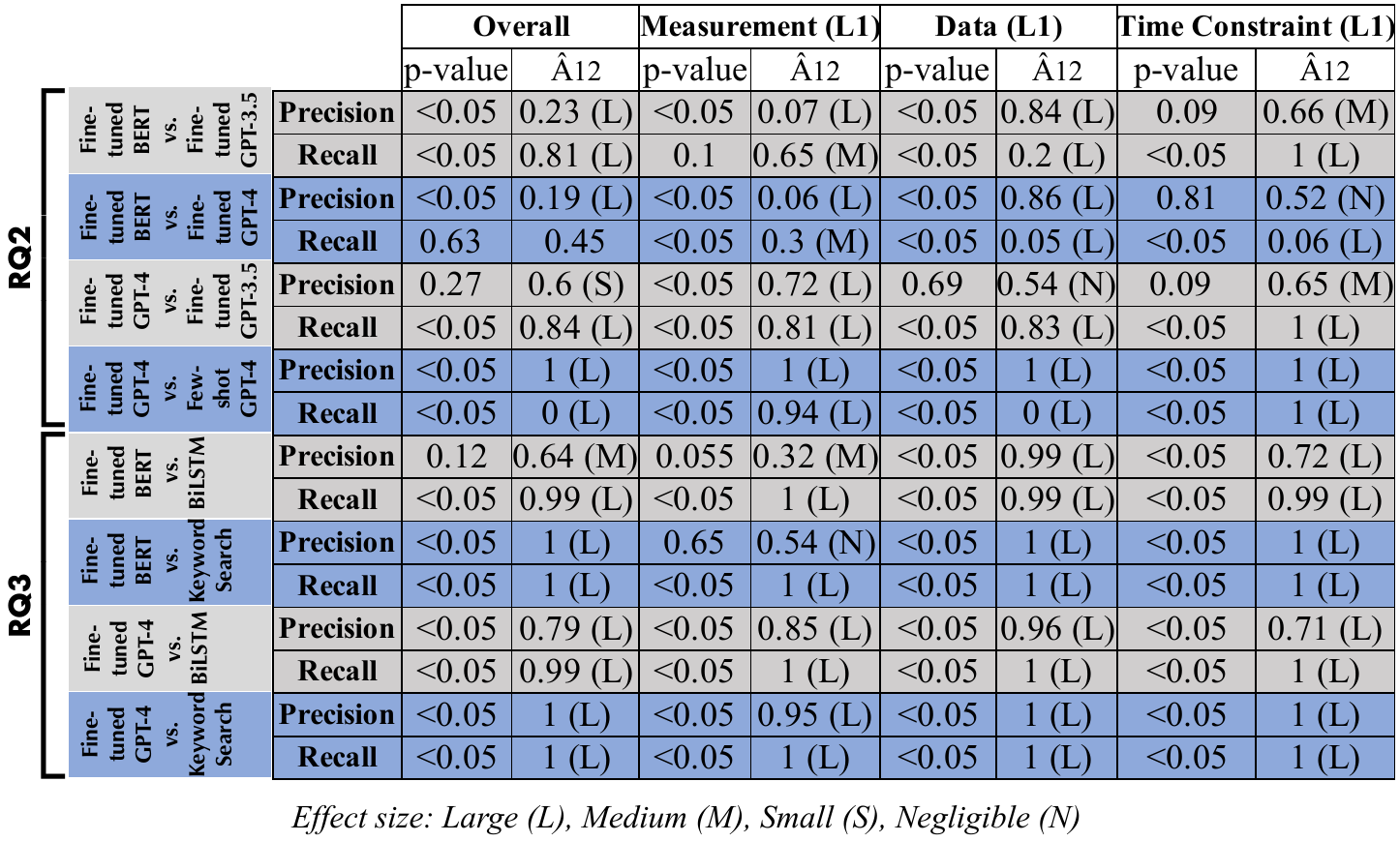}
\end{table}

\sectopic{Accuracy statistics for BERT base, GPT-3.5 and GPT-4.}
Figures~\ref{fig:Precision} and \ref{fig:Recall} present the average accuracy metrics from 20 runs for the fine-tuned BERT base, fine-tuned GPT-3.5-turbo, fine-tuned GPT-4o, and GPT-4o with few-shot learning. These results were obtained following the evaluation procedure explained in EXPI (Section~\ref{subsec:evaluationprocedure}). In Table~\ref{tab:Statistical Significance Testing}, we statistically compare the accuracy of the most practically relevant pairs selected from these four models for both the \emph{Overall} concept and the level-1 (L1) concepts, as shown in Fig.~\ref{fig:metadata}. We do so using the Wilcoxon Rank-Sum Test and the Vargha-Delaney effect size, $\hat{A}_{12}$, as described in Section~\ref{subsec:metricS}.
The p-values in Table~\ref{tab:Statistical Significance Testing} assess whether the difference between two models is statistically significant. A p-value less than 0.05 (i.e., $p < 0.05$) suggests that the observed accuracy difference between the models is statistically significant, meaning it is unlikely to have occurred by chance. To interpret the $\hat{A}_{12}$ values, note that in a comparison between \emph{Model}$_A$ and \emph{Model}$_B$, i.e., ``\emph{Model}$_A$ vs. \emph{Model}$_B$'', the value of $\hat{A}_{12}$ is directional: $\hat{A}_{12} > 0.5$ favours \emph{Model}$_A$, while $\hat{A}_{12} < 0.5$ favours \emph{Model}$_B$.

\sectopic{Comparing the fine-tuned models.} Starting with the three fine-tuned models, we observe from the results for the \emph{Overall} concept that, in terms of \emph{Precision}, fine-tuned GPT-3.5-turbo and fine-tuned GPT-4o significantly outperform fine-tuned BERT base, with averages of $\approx$89\% each, compared to fine-tuned BERT's $\approx$87\%. For \emph{Recall}, fine-tuned BERT base and fine-tuned GPT-4o significantly outperform fine-tuned GPT-3.5-turbo, with averages of $\approx$87\% and 86\%, respectively, compared to fine-tuned GPT-3.5-turbo's $\approx$83\%.

Based on the above, fine-tuned GPT-4o is a better alternative to fine-tuned GPT-3.5-turbo for the \emph{Overall} concept in terms of both \emph{Precision} and \emph{Recall}. This advantage extends to the more specific L1 concepts of \emph{Measurement}, \emph{Data}, and \emph{Time Constraint}. As shown in Figs.~\ref{fig:Precision} and \ref{fig:Recall}, and Table~\ref{tab:Statistical Significance Testing}, fine-tuned GPT-4o significantly outperforms fine-tuned GPT-3.5-turbo in \emph{Recall} across all L1 concepts. For \emph{Precision}, fine-tuned GPT-4o has a statistically significant advantage over fine-tuned GPT-3.5-turbo in \emph{Measurement}. In the cases of \emph{Time Constraint} and \emph{Data}, fine-tuned GPT-4o's average \emph{Precision} is either equal to or slightly better than GPT-3.5-turbo’s, though these differences are not statistically significant. 
Similar trends are observed for the L2 concepts, with statistical results provided in our online material for brevity~\citep{EvaluationResults}.
Taken together, our findings indicate that when models are fine-tuned, GPT-4o offers an advantage over GPT-3.5-turbo for our classification task.

With fine-tuned GPT-4o shown to outperform fine-tuned GPT-3.5-turbo, the next question is how it compares to fine-tuned BERT base. As stated earlier, for the \emph{Overall} concept, fine-tuned GPT-4o significantly outperforms fine-tuned BERT base in \emph{Precision}, though neither model shows a statistically significant advantage in \emph{Recall}. At the L1 concept level, fine-tuned GPT-4o significantly outperforms fine-tuned BERT base in \emph{Precision} and \emph{Recall} for \emph{Measurement}, \emph{Recall} for \emph{Data}, and \emph{Recall} for \emph{Time Constraint}. Fine-tuned BERT base, on the other hand, has significantly higher \emph{Precision} for \emph{Data}. No other significant differences were observed. 

Since fine-tuned BERT base never outperforms fine-tuned GPT-4o in \emph{Recall}; its higher \emph{Precision} for \emph{Data} is offset by lower \emph{Recall}; and similar trends are observed for L2 concepts (with statistical results provided in our online material for brevity~\citep{EvaluationResults}), fine-tuned GPT-4o presents the better alternative in comparison with fine-tuned BERT base as well.

In summary, considering that fine-tuned GPT-4o presents a better alternative than both fine-tuned GPT-3.5-turbo and fine-tuned BERT base without any major trade-offs, we conclude that GPT-4o is the most suitable model for our classification task in the fine-tuning scenario.

\sectopic{Comparing GPT-4o fine-tuning and few-shot learning.} 
As shown in Figs.~\ref{fig:Precision} and \ref{fig:Recall}, for the \emph{Overall} concept, GPT-4o with few-shot learning achieves an average \emph{Precision} of $\approx$65\%, about 24\% lower than fine-tuned GPT-4o. However, GPT-4o with few-shot learning achieves a notably high \emph{Recall} of $\approx$97\%, outperforming fine-tuned GPT-4o by about 10\%. This trend largely extends to the L1 concepts: fine-tuned GPT-4o shows higher \emph{Precision} across all three L1 concepts, while GPT-4o with few-shot learning achieves better \emph{Recall} for two (\emph{Data} and \emph{Time Constraint}), but lags by 5\% over \emph{Measurement} ($\approx$81\% vs. $\approx$86\%). All differences are statistically significant with large effect sizes, as shown in Table~\ref{tab:Statistical Significance Testing}.

The above results highlight an important trade-off: For our classification task, while fine-tuning improves \emph{Precision}, it generally has a negative impact on  \emph{Recall}. Interestingly, all the examples used in our few-shot learning experiment are drawn from our fine-tuning/training set, meaning that our fine-tuned models are exposed to all the few-shot examples (along with many others).  The phenomenon observed here aligns with findings in the literature on LLMs, which suggest that providing a large number of specific examples during fine-tuning may lead to overfitting, causing the fine-tuned model to pick up on spurious correlations and also potentially reducing its generalization abilities~\citep{mosbach2023few}.

From a practical standpoint, fine-tuning a powerful model like GPT-4o for our classification task has significantly improved \emph{Precision}, but, broadly speaking, has also reduced \emph{Recall} considerably. Although the gain in \emph{Precision} observed in our experiments is more than twice the loss in \emph{Recall}, the latter often holds greater importance in requirements engineering, where filtering out false positives is generally an acceptable trade-off if it helps to avoid missing relevant labels. Another crucial factor to consider with fine-tuning is the cost of creating a sufficiently representative fine-tuning dataset. In real-world applications, this process can be costly, which may influence the decision on whether fine-tuning is worthwhile. 

The above said, it is critical to note that our few-shot scenario represents a ``best case'' for few-shot learning. This is because we benefited from hindsight gained during our GT study, with the luxury of reflecting on numerous examples from that study and selecting a handful of the most comprehensive and inclusive ones for our few-shot prompt, ensuring the broadest possible coverage of the concepts. The question remains whether, in a domain where analysts are less familiar with the subject matter and lack the ability to choose from a large pool of examples as we did, they can still identify effective examples for few-shot learning that capture all the key aspects. Our current empirical evidence does not provide an answer to this question, as addressing it would require a separate line of investigation, such as user studies or controlled experiments using examples provided by non-authors for few-shot learning.

\begin{table}
\vspace*{.5em}
\caption{Accuracy Results for Scarce Concepts.}
\label{tab:scarce}
\centering
\includegraphics[width=0.32\linewidth]{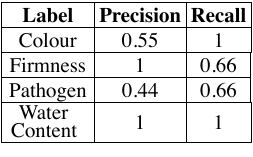}
\end{table}
\sectopic{Handling scarce concepts.}
As mentioned in Section~\ref{subsec:dataset}, certain concepts in our content model (Fig.~\ref{fig:metadata}), specifically \emph{Colour}, \emph{Firmness}, \emph{Pathogen}, and \emph{Water Content}, are scarce. For these concepts, BERT performs poorly, with both \emph{Precision} and \emph{Recall} falling below 5\%. GPT-3.5 and GPT-4 show progressively better results than BERT; however, accuracy still remains low (detailed results can be found in our online material~\citep{EvaluationResults}). To improve accuracy for scarce concepts, we employ keyword-based classification, as discussed in Section~\ref{subsec:keyword-based}. Since keyword-based classification introduces no random variation, we have excluded the results for these scarce concepts from the plots in Figs.~\ref{fig:Precision} and \ref{fig:Recall}. Instead, the results for the scarce concepts are presented in Table~\ref{tab:scarce}.

\sectopic{Comparing accuracy over Canadian vs. US regulations.} Our LLM-based classification pipeline has comparable accuracy for both Canadian (FSRG) and US (FDA) regulations. In Table~\ref{tab:FDA-nonFDA}, we present the average (Avg) and standard deviation (SD) of accuracy metrics, calculated over 20 runs of the pipeline when instantiated with (fine-tuned) BERT, GPT-3.5, or GPT-4. The non-parametric pairwise Wilcoxon rank-sum test, performed on the results of the individual LLMs, indicates no statistically significant difference between FSRG and FDA in terms of \emph{Recall} (at the 5\% confidence level). For \emph{Precision}, there is no statistically significant difference in BERT results, but GPT-3.5 and GPT-4 perform statistically significantly better over FDA. These findings suggest that our pipeline performs equally well, if not better, on US regulations, despite the customization data being exclusively derived from Canadian regulations. Further details on our statistical tests can be found in our online material~\citep{EvaluationResults}.

\begin{table}
\caption{Accuracy for Canadian (FSRG) vs. US (FDA) Regulations.}
\label{tab:FDA-nonFDA}
\centering
\includegraphics[width=0.4\linewidth]{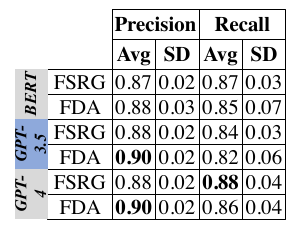}
\end{table}

\begin{samepage}
\begin{mdframed}[style=MyFrame]
\emph{The answer to \textbf{RQ2} is: 
Among fine-tuned models, GPT-4 outperformed GPT-3.5 and BERT, achieving $\approx$89\% \emph{Precision} and $\approx$87\% \emph{Recall} for our classification task. Comparing GPT-4 with few-shot learning against its fine-tuned version revealed a trade-off: few-shot learning increased \emph{Recall} to $\approx$97\% but decreased \emph{Precision} to $\approx$65\%.  
Our few-shot learning results may nonetheless be overoptimistic due to carefully selected examples. 
Despite being developed exclusively with Canadian regulations, our classification pipeline achieved comparable or better accuracy on selected US regulations, suggesting a degree of generalization across regulatory jurisdictions.
}
\end{mdframed}
\end{samepage}

\subsubsection{RQ3: How do LLM-based classification models compare to simpler baselines in the context of food-safety regulations?} \label{subsec:RQ3}

\begin{figure*}
\centering\mbox{\hspace*{-2em}}
    \includegraphics[width=1.044\linewidth]{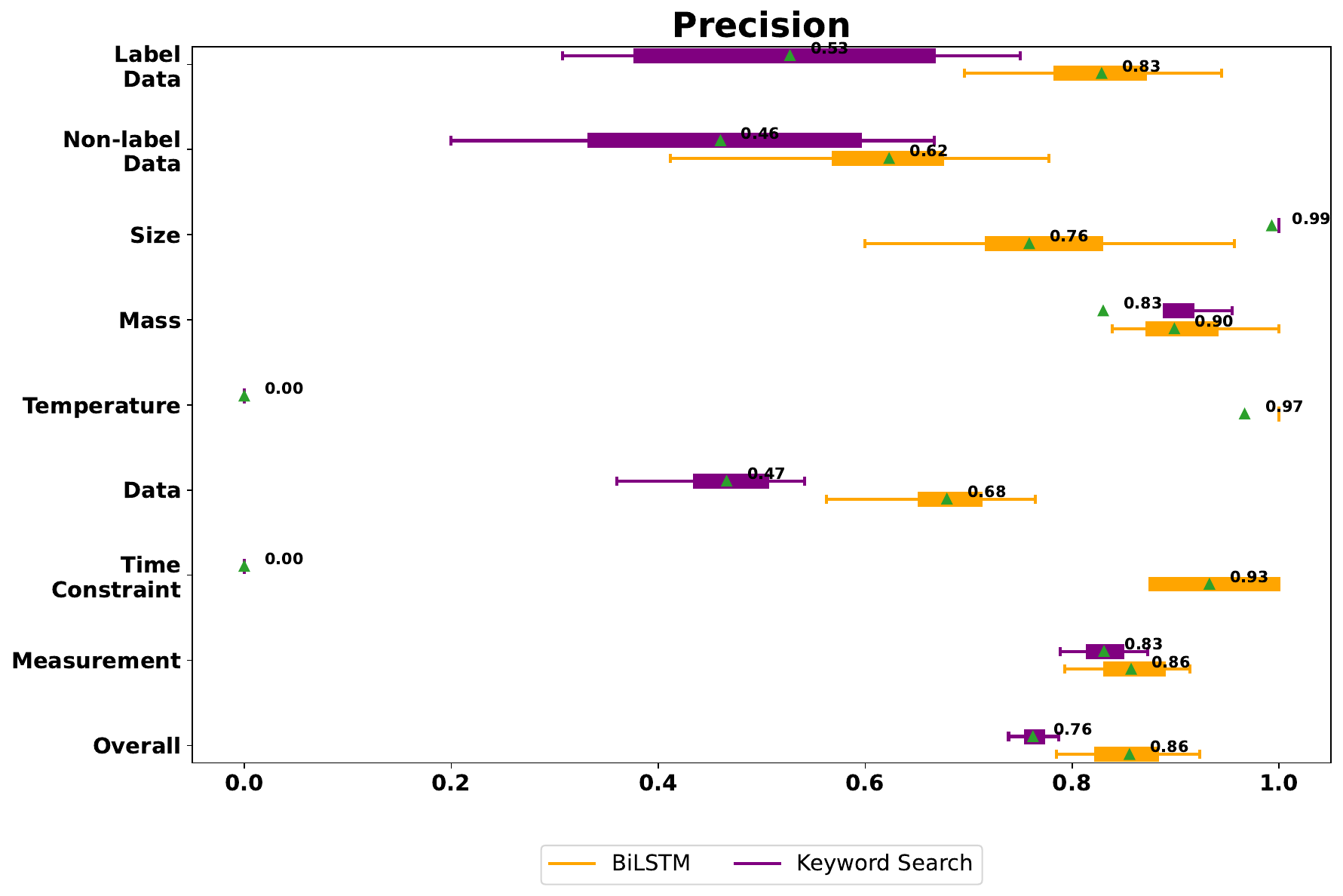}
\caption{\emph{Precision} for BiLSTM and Keyword Search (the $x$-axis represents the \emph{Precision} values).}
\label{fig:Precision baselines}

\centering\mbox{\hspace*{-2em}}
    \includegraphics[width=1.044\linewidth]{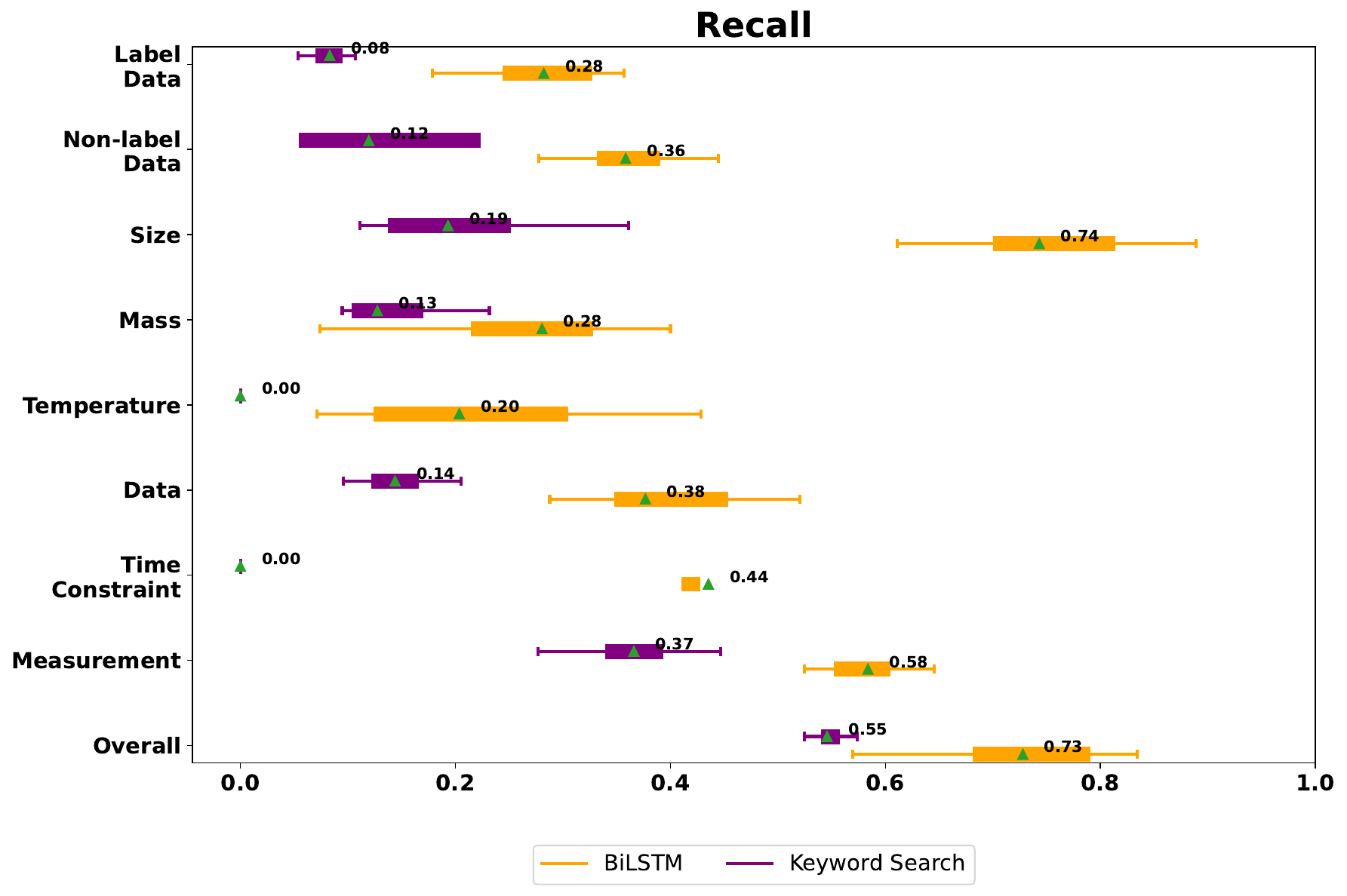}
\caption{\emph{Recall} for BiLSTM and Keyword Search (the $x$-axis represents the \emph{Recall} values).}
\label{fig:Recall baselines}
\end{figure*}

Figures~\ref{fig:Precision baselines} and~\ref{fig:Recall baselines} present boxplot results for 20 runs of the BiLSTM and keyword-search baselines.
In the last four rows of Table~\ref{tab:Statistical Significance Testing} (shown earlier), we compare the accuracy of our best-performing models from RQ2 with that of these baselines.

The results indicate that BERT base and GPT-4o consistently outperform the baselines with large effect sizes in terms of \emph{Recall}.

As for \emph{Precision}, with one exception, BERT base either outperforms the baselines with statistical significance or achieves better results without statistical significance on the \emph{Overall} and all L1 concepts. The exception is the \emph{Measurement} concept, where BiLSTM yields a $\approx$3\% better average \emph{Precision}, albeit without statistical significance. For \emph{Measurement}, BERT's \emph{Recall} is $\approx$25\% better. The 3\% deficit in \emph{Precision} is thus a reasonable trade-off for the much better \emph{Recall}. 
 The statistical tests for L2 concepts, provided online \citep{EvaluationResults}, reveal a similar trend: BERT base is consistently better in terms of \emph{Recall} and never at a major disadvantage in terms of \emph{Precision}, given its considerably better \emph{Recall}.

In a similar vein, GPT-4o consistently outperforms the baselines with large effect sizes in terms of \emph{Precision} on the \emph{Overall} and across all L1 concepts. The statistical significance tests for L2 concepts, provided online \citep{EvaluationResults}, reveal a similar trend with two exceptions: \emph{Label Data} has a \emph{Precision} of $\approx$79\% (GPT-4o) vs. $\approx$83\% (BiLSTM), and \emph{Temperature} has a \emph{Precision} of $\approx$87\% (GPT-4o) vs. $\approx$97\% (BiLSTM), with the difference in \emph{Temperature} being statistically significant. However, the statistically significant gains in \emph{Recall} -- $\approx$41\% and $\approx$69\%, respectively, for these two concepts -- favour GPT-4o, making it the better choice despite the (minor) deficits in \emph{Precision}.

\vspace*{.5em}
\begin{samepage}
\begin{mdframed}[style=MyFrame]
\emph{The answer to {\bf RQ3} is: For our classification task, BERT and GPT consistently outperform simpler, non-Transformer baselines in terms of \emph{Recall}. Moreover, BERT and GPT have no significant disadvantage in \emph{Precision}, with any \emph{Precision} loss offset by much larger gains in \emph{Recall}.}
\end{mdframed}
\end{samepage}

\subsection{Limitations and Validity Considerations} \label{sec:threats}
In this section, we discuss limitations and threats to validity. The validity considerations most pertinent to our empirical evaluation are \emph{internal validity}, \emph{external validity}, \emph{construct validity}, and \emph{statistical conclusion validity}.

\sectopic{Limitations.}
An important limitation of our evaluation is that our classification pipeline has not yet been applied in an industrial context. While, as mentioned in Section~\ref{sec:introduction}, our research was driven by a need identified by an industry partner, we have not yet reached the stage of deploying our pipeline in a full-scale industrial use case, such as for formulating the regulatory requirements of an actual food-safety monitoring system. As a result, we lack empirical evidence to evaluate the potential benefits of automating the classification of food-safety regulations in the specific industrial context that motivated our work.

\sectopic{Internal validity.} The main threat to internal validity is bias. We implemented several measures to mitigate bias risks. First, two authors jointly examined all provisions in our fine-tuning/training set and discussed challenging cases. Second, we delegated the task of annotating our test data entirely to third-party annotators (non-authors), who had no knowledge of our automated classification solution. 
Finally, we measured inter-rater agreement over 10\% of the test data to assess the reliability of the third-party annotators' work.

\sectopic{External validity.}
Our evaluation focuses exclusively on food-safety regulations in Canada, a specific legal domain within one country. Consequently, our findings may have limited broader applicability. However, it is important to recognize that food safety is one of the most far-reaching regulatory frameworks, given its direct impact on public health and well-being worldwide. Furthermore, as noted in Section~\ref{sec:Back}, Canada ranks highly for ``Quality and Safety'' in the Global Food Security Index, indicating that its food-safety regulations provide a good foundation for studying this area.

Another key consideration here is that regulatory requirements analysis has traditionally been domain-specific, as this approach yields more focused and concrete outcomes by addressing challenges unique to each domain~\citep{Breaux2006Towards,Ghanavati2007Towards,Sleimi2018Automated}. We follow this practice, but acknowledge that it inherently imposes limitations on generalizability. While further case studies remain essential, we observe that our classification pipeline consistently performs well on test data from both Canada and the US. This finding lends confidence to the external validity of our results, especially within comparable regulatory environments.

\sectopic{Construct validity.}
Our quantitative analysis of classification accuracy followed established best practices, employing the standard metrics of \emph{Precision} and \emph{Recall}. As explained in Section~\ref{sec:extraction}, we chose individual sentences as the units of classification. This choice is supported by existing research on the classification of legal texts and considering the structural and conceptual parallels between food-safety regulations and other regulatory frameworks, such as data protection and privacy laws, examined in prior studies.

\sectopic{Statistical conclusion validity.}
To compare the accuracy of different LLMs, we used the Wilcoxon Rank-Sum Test. This non-parametric test enables us to compare the distributions of two independent samples -- in our context, the distributions of \emph{Precision} or \emph{Recall} values from pairs of models, e.g., BERT base and GPT-4o. This test is well-suited to our analysis because our data -- \emph{Precision} or \emph{Recall} values from 20 runs of each alternative solution -- cannot be assumed to follow a normal distribution. Our use of the Wilcoxon Rank-Sum Test aligns with the guidelines provided by \citet{arcuri2011practical} for evaluating randomized algorithms in software engineering. We note that when comparing multiple alternatives -- in our context, multiple LLMs -- some researchers recommend first applying the Kruskal–Wallis Test to determine whether any overall significant differences exist among the alternatives before proceeding with Wilcoxon Rank-Sum Tests between pairs of alternatives. However, because we only had a small number of pairwise comparisons in our work, and we were specifically interested in direct comparisons between pairs of alternatives, we opted to bypass the Kruskal–Wallis Test and proceed directly with Wilcoxon Rank-Sum Tests. Conducting multiple statistical tests is known to increase the risk of Type I error inflation. Nonetheless, in accordance with the recommendations of \citet{arcuri2011practical}, we have chosen not to apply any corrections (e.g., Bonferroni correction) to control the family-wise error rate. This decision is particularly justified given the relatively small number of pairwise comparisons in our study.

We considered alternative tests, such as the Wilcoxon Signed-Rank Test, McNemar's Test, and the 5x2 cross-validation F-test, but favoured the Wilcoxon Rank-Sum Test for methodological reasons.
The Wilcoxon Signed-Rank Test requires \emph{multiple} datasets. Since we have only a single dataset (test set), applying this statistical test would have required splitting our dataset into smaller subsets. We concluded that this approach would not be methodologically sound, as it would artificially create multiple datasets from a single one, which was collected within the same domain. McNemar's Test also proved unsuitable, as it is primarily designed for nominal data and does not allow for direct comparison of our metrics, which are numeric. Finally, the 5x2 cross-validation F-test was ruled out to avoid introducing bias through the mixing of training and test sets. Given the effort put into constructing a separate test set with third-party annotators, mixing this data with the training data -- noting that the training data was author-annotated -- would have conflicted with our objective of mitigating potential author bias.

Despite these precautions, we acknowledge that relying on a single test set limits the robustness of our statistical findings. Future studies with multiple test sets from diverse domains are needed to further increase robustness.

\section{Discussion} \label{sec:discussion}
This section provides practical insights into the implications of the empirical results presented in Section~\ref{sec:evaluation}.

\sectopic{In our application context, LLMs tend to outperform simpler models.} 
Our comparison with baseline models shows that, for our classification task, BERT and GPT are clearly more accurate than simpler models, such as LSTM and feature-based learning (specifically, Random Forest) when applied to keywords. This increased accuracy can be attributed to several factors, including the LLMs' ability to better understand context and meaning through pre-training, their stronger capacity to generalize across tasks, and their attention mechanisms, which help manage long-range dependencies more effectively than models like LSTM. 

While it is generally unsurprising that LLMs outperform simpler models, providing  evidence to justify their use for a specific task is still important. LLMs require considerably more computational power for training, fine-tuning, and querying, resulting in higher costs and a greater environmental impact. Therefore, striking a balance between accuracy and complexity -- following the ``Goldilocks principle'' -- is key to ensuring that the selected model is neither too simple nor too complex, but optimally suited \hbox{for the task at hand.}

Furthermore, it is important to recognize that not all LLMs or their variants perform equally well in every context. In our experiments, for instance, the specific Llama and Mixtral variants we tested did not consistently follow instructions, resulting in unpredictable outputs. This unpredictability made it challenging to use these models effectively, and consequently, conducting a systematic evaluation of their accuracy was not feasible.

\sectopic{Using few-shot learning vs. fine-tuning could imply a trade-off between \emph{Precision} and \emph{Recall}.} 
Emerging evidence suggests that out-of-the-box LLMs, unless customized to some extent, face challenges with domain-specific software engineering tasks~\citep{ahmed2024can}. In this article, we addressed one such task by focusing on food-safety regulations and adapting GT to assign meanings to concept labels. To classify accurately according to these labels, we needed to customize the LLM to capture the nuanced meanings and subtleties of the concepts identified in our GT study.

To customize an LLM, one can select between two main options: fine-tuning and few-shot learning, both of which were considered in RQ2 for GPT-4o. Our findings indicate that these two options lead to different trade-offs. Fine-tuning, which specializes a model by adjusting its weights based on task-specific data, guides the model towards a detailed interpretation of concepts but potentially at the expense of narrowing its general understanding and making it overly confident in specific interpretations. Consequently, the model may overlook other valid variations and dismiss relevant instances that do not precisely align with the fine-tuning data, leading to a reduction in \emph{Recall}. Few-shot learning, in contrast, maintains broader adaptability by presenting the LLM with only a few representative examples. This can result in  more general understanding capabilities and, in turn, increase \emph{Recall}. However, the flexibility may come at the expense of \emph{Precision}, as few-shot learning may not provide enough guidance for the LLM to focus on important details necessary for accurately recognizing subtleties.

Resolving the above trade-off ultimately depends on whether the task prioritizes \emph{Precision} or \emph{Recall}. Although this trade-off may be inevitable, caution is needed when interpreting our current few-shot learning results. Specifically, based on our existing empirical data, we cannot determine whether comparable \emph{Precision} levels can be achieved in a few-shot learning scenario where the provided examples are not as optimal. Nor can we ascertain whether adding more examples in few-shot learning could further increase \emph{Precision} without compromising \emph{Recall}. Addressing these questions requires substantial additional experimentation. However, based on our existing evidence, we can state that, for models like GPT-4o with extensive pre-training knowledge, few-shot learning shows promise as a more efficient and cost-effective alternative to fine-tuning for our classification task.

\sectopic{There is more to generative LLMs than classification accuracy.} 
In our experiments with fine-tuned models, although GPT-4o slightly outperforms BERT, the practical advantage is marginal: GPT-4o shows only a 1\% improvement in \emph{Recall} (87\% vs. 86\%) and a 2\% improvement in \emph{Precision} (89\% vs. 87\%) on average. These results were obtained under identical fine-tuning conditions; thus, the effort spent on curating fine-tuning data was not a deciding factor. We conclude, therefore, that for our specific classification task, a fine-tuned BERT model performs nearly as accurately as a fine-tuned GPT model. Given BERT's significantly smaller computational footprint, the question arises whether using a resource-intensive model like GPT is justified for our classification task.

Our answer to this question is as follows: Classification is often an important but preliminary task for many follow-on tasks that analysts need to perform. Generative models like GPT offer additional capabilities, such as providing explanations, rationale, and the ability to follow conversational instructions. These features are key for stakeholders who require justification for decisions or wish to engage interactively in the classification process, correcting mistakes as they arise. The ability to request explanations or provide real-time feedback is absent in non-generative models like BERT. In our application context, the added value of generative models like GPT lies not necessarily in a drastic improvement in classification accuracy but in supporting more natural, human-in-the-loop interactions with legal texts in the future.

\sectopic{
The classification pipeline's execution time is sufficient for practical use.} We measured the execution times of our classification pipeline using different LLMs. Since our classification task is a one-time process, execution times do not impact the choice of which LLM to use, as long as the times remain reasonable. Below, we discuss the execution times of our most accurate pipelines, which were created using BERT and GPT, noting that these times are not directly comparable because the models were deployed in different environments.

To measure BERT's execution time, we used a modest platform to replicate the resources available to end-users. Specifically, we used the free version of Google Colab Cloud with the following specifications: Intel Xeon CPU@2.30GHz, Tesla T4 GPU, and 13GB RAM. 
BERT's execution time includes a one-time pre-processing and a one-time loading phase, followed by label predictions in batches of 16. In our Colab setup, pre-processing and loading together took $<$\,0.5 seconds. Label prediction with BERT base for one batch took $\approx$4.32 seconds, resulting in $\approx$0.27 seconds per provision. 
For GPT, we used OpenAI's token-based plan (paid subscription service). Our GPT experiments employed the same pre-processing as BERT, with the loading of the (fine-tuned) LLM managed by OpenAI. Calculating precise execution times for GPT is not feasible due to lack of control over OpenAI loads during execution as well as factors such as rate limits and OpenAI's safeguards against disruptive attacks. For both GPT-3.5-turbo and GPT-4o, the average time to complete one run over our entire test set was $\approx$5 minutes, yielding an average execution time of $\approx$0.7 seconds per provision. These results indicate that execution time is not a barrier to adoption, irrespective of the LLM being used.

\section{Related Work} \label{sec:related}
We discuss previous work on~(1) extracting metadata from legal texts, (2)~applying LLMs to natural-language requirements, and~(3) monitoring food safety through IoT. To our knowledge, no prior work exists on automated classification of requirements-related provisions in food-safety regulations.

\subsection{Metadata Extraction from Legal Texts}
Legal documents are typically lengthy and complex, making it difficult for individuals to locate relevant information quickly and accurately. This has led to the development of automated approaches for information extraction from legal texts.

\citet{Breaux2006Towards} as well as \citet{Breaux2008Analyzing} use semantic parameterization to extract rights and obligations from regulatory texts, and further propose an approach for traceability during requirement extraction \citep{Breaux2013Preserving}. \citet{Bhatia2018Semantic} address privacy-policy incompleteness by representing data actions as semantic frames. These works, despite being amongst the first to focus on information extraction from legal texts, do not use ML techniques and are limited to privacy policies.

\citet{Amaral2022AI} employ a combination of NLP and feature-based learning to extract metadata from privacy policies and assess their compliance with GDPR. In subsequent research \citep{amaral2023nlp,amaral2023ml,ilyas2023multi}, the approach is expanded to incorporate metadata related to GDPR's data processing agreements (DPAs). These prior studies differ from ours in both domain and the utilization of feature-based learning as opposed to LLM-based approaches.

Some previous studies link privacy policies to software code. \citet{Fan2020Empirical} use traditional ML classifiers to check Android mobile health app compliance with GDPR, while \citet{Hamdani2021Combined} combine ML and rules for compliance checking of privacy policies with GDPR. \citet{Xie2022Scrutinizing} employ Bayesian classifiers and NLP to check whether Amazon Alexa's skills meet their privacy requirements. In contrast, our approach focuses on food-safety regulations. We systematically evaluate models from two families of LLMs -- BERT and GPT -- for the classification of these regulations.

\citet{Zeni2015Gaiust,Zeni2016Building} and \citet{Sleimi2018Automated} extract semantic metadata from legal texts using traditional NLP and semantic web techniques. \citet{Abualhaija2022Automated} and \citet{Abualhaija2024AI} employ LLMs for legal question answering and the identification of regulatory changes, respectively. And, \citet{Hassani2024Rethinking} use LLMs for legal compliance checking. These studies differ from ours in terms of analytical goals, inputs, and the application of NLP and LLM techniques.

\subsection{LLMs and Natural-language (NL) Requirements}
Several studies use LLMs for analyzing NL requirements, with tasks including classifying non-functional requirements \citep{Hey2020NoRBERT,Chatterjee2021Pipeline,Alhoshan2022Zero-Shot}, classifying and summarizing contractual obligations \citep{Sainani2020Extracting,jain2023transformer}, detecting requirements smells \citep{Habib2021Detecting}, classifying security requirements \citep{Varenov2021Security}, classifying user feedback \citep{Mekala2021Classifying}, classifying requirements dependencies \citep{Deshpande2021BERT}, detecting causality \citep{Fischbach2021Automatic}, classifying and clustering coreferences \citep{Wang2020Deep,Wang2022Detecting}, transforming NL requirements into formal specifications \citep{Nayak2022Req2Spec}, predicting issue links \citep{Luders2022Automated}, classifying issue sentences \citep{Mehder2022Classification}, identifying similar requirement \citep{Abbas2022Relationship}, checking completeness \citep{Luitel2023Using}, generating elicitation-interview scripts \citep{Gorer23Generating}, named entity recognition in software specification documents \citep{das2023zero}, detecting variability in NL requirements documents \citep{fantechi2024exploring}, formal requirements engineering \citep{spoletini2024return},  requirements quality assurance~\citep{Fazelnia2024Lessons}, and checking requirements satisfiability~\citep{Santons2024InContext}. These studies demonstrate the versatility of LLMs in requirements-analysis tasks. We have benefited from the best practices in the above-cited strands of work; however, our analytical objectives set us apart.

\subsection{Food-monitoring Systems}
Food monitoring is a crucial area of research for improving food safety and enhancing supply-chain efficiency. \citet{Bouzembrak2019Internet} have conducted a systematic literature review to examine the potential of IoT for food safety and quality monitoring, as well as food traceability and supply chain. They observe that most existing IoT research aimed at these objectives focuses on measurement solutions using sensors and communication technologies. These solutions are technology-driven and not directly linked or traceable to regulations. Our GT study highlights, from a regulatory perspective, the importance of all the core measurement types identified in the literature. The close alignment between the measurements implied by food-safety regulations and the existing measurement technologies confirms the validity of our interpretation of systems/software relevance in the context of food safety. Upon snowballing from the above literature review, we found no research that focuses on legal text processing or on establishing traceability between measurements and regulatory requirements. 

\citet{brewer2021trust} introduce a trust framework, and \citet{pearson2021food} deploy food data trusts, both to facilitate secure information sharing across food-supply chains, aiming for regulatory compliance and enhancing transparency. \citet{pearson2023decarbonising} further apply digital technologies to decarbonize food systems, focusing on supply-chain transparency and sustainability. These studies align closely with our work's emphasis on the potential of IoT for ensuring food safety. Yet, the studies do not address automated processing of food-safety regulations.

The closest work to ours is by \citet{Markovic2016Modelling}, who propose an ontological model for recording provenance in the food-safety domain and demonstrate its utility in the context of automated provenance generation for food-safety compliance checking. They discuss measurements but limit themselves to temperature, whereas our content model (Fig.~\ref{fig:metadata}) is far richer.

\section{Conclusion} \label{sec:conclusion}
We presented two main contributions: (1) a Grounded Theory study that characterizes food-safety concepts in regulatory provisions impacting modern software-intensive food-safety systems, and (2) an LLM-based classification pipeline that classifies the provisions of food-safety regulations based on their relevance to systems and software requirements. We examined the effectiveness of our LLM-based classification pipeline by instantiating it with two families of LLMs: BERT and GPT. 
Fine-tuning GPT-4o yielded the highest accuracy, achieving 89\% \emph{Precision} and 87\% \emph{Recall}. When we implemented few-shot learning with GPT-4o, \emph{Recall} improved to 97\%, but \emph{Precision} dropped to 65\%. While our classification pipeline was tailored to Canadian regulations, it effectively processed provisions from both Canadian and US regulations. Finally, we observed that LLMs significantly outperformed simpler, non-Transformer-based baselines.

Although our experimental results do not directly measure practical usefulness, they indicate that the best-performing LLMs achieve high \emph{Precision} and \emph{Recall}, suggesting that automation is valuable for our classification task. Manual review of regulatory documents is known to be tedious and error-prone, especially under the time pressures common in industrial environments. In this context, even less-than-perfect automated results can be beneficial, provided that \emph{Recall} remains consistently high -- an outcome observed with the most accurate models in our evaluation. 

As part of our work, we curated a dataset containing over a thousand labelled provisions. We are sharing this dataset with the research community to stimulate further research in the domain of food supply chain and safety. In future work, we plan to (1)~study the direct derivation of requirements from food-safety regulations, (2)~explore the generalizability of our food-safety concepts beyond North America, and (3)~conduct user studies to evaluate the practical benefits of our automated classification solution.

\section{Data Availability}\label{sec:package}
Our complete replication package is available on both GitHub \citep{Replication} and Zenodo \citep{Zenodo}. The package includes:
\begin{itemize}
    \item Our dataset, keywords, and SFCR exclusions \citep{Data};
    \item The hyperparameters and dataframes for different models, along with complementary statistical analysis results \citep{EvaluationResults};
    \item The implementation of all algorithms, including baselines and evaluation scripts \citep{Code}.
\end{itemize}

\section*{Compliance with Ethical Standards}
\sectopic{Conflict of Interest.} The authors declare that they have no conflicts of interest.

\sectopic{Acknowledgments.} We are grateful to the engineering team at Stratosfy, especially Madan Kanala, for stimulating discussions.

\sectopic{Funding.} This research was supported by the Natural Sciences and Engineering Research Council of Canada (NSERC) through the Discovery, Discovery Accelerator, and Alliance programs, by the Ontario Centre of Innovation (OCI) through the VIP program, and by Stratosfy.

\sectopic{Ethical Approval.} This research did not involve human participants or animals; therefore, ethical approval was not required.

\sectopic{Informed Consent.} No personal data or identifiable information is reported in this research; informed consent is not applicable.

\sectopic{Author Contributions.} This article is part of the PhD studies of the first author, who has led the work, including conceptualization, methodology, analysis, and writing. The second and third authors supervised the research, contributed to the directions and design, and provided substantial input into the writing and interpretation of the results.

\bibliography{ref}
\end{document}